\documentclass[hyper]{JHEP3}
\usepackage{amsmath,amssymb}

\usepackage{ifpdf}
\ifpdf
  \usepackage[pdftex]{graphicx}
\else
  \usepackage[dvipdfm]{graphicx}
\fi

\title{Wall-crossing, Toric divisor and Seiberg duality}
\author{Takahiro Nishinaka\\
NHETC and Department of Physics and Astronomy, Rutgers University,\\
Piscataway, NJ 08854, USA\\
\\
{\tt nishinaka@physics.rutgers.edu}}
\abstract{
We study the wall-crossing phenomena of BPS D4-D2-D0 states on the conifold and orbifold $\mathbb{C}^2/\mathbb{Z}_2$, from the viewpoint of the quiver quantum mechanics on the D-branes. The K\"ahler moduli dependence of the BPS index is translated into the FI parameter dependence of the Witten index. The wall-crossing phenomena are related to the Seiberg dualities of the quiver quantum mechanics. All the differences from the D6-D2-D0 case arise from the additional superpotential and ``anti-quark'' induced by the D4-brane. 
When the D-branes are on the conifold, the flop transition changes the duality cascade. When the D-branes are on the orbifold $\mathbb{C}^2/\mathbb{Z}_2$, the generating function of the Witten index is always given by a character of the affine SU(2) algebra. Both are consistent with the wall-crossing formula for BPS indices.
}
\preprint{
RUNHETC-2013-09
}

\begin{document}

\bibliographystyle{JHEP}   

%%%%%%%%%%%%%%%%%%%%%%%%%%%%%%%%%%%%%%%%%%

\section{Introduction}

The spectrum of D-brane bound states captures various non-perturbative physics in string theory. In particular, in type IIA string theory on a Calabi-Yau three-fold, the spectrum receives world-sheet instanton corrections controlled by $\alpha'/z$,
where $z$ is the K\"ahler moduli of the background Calabi-Yau geometry. In the large radius limit $|{\rm Im}\,z|\to \infty$, the bound state of D-branes is described in the field theory on heavier branes, in which lighter branes are realized as instantons and magnetic fluxes. On the other hand, in the small radius limit ${\rm Im}\,z\to 0$, the D-brane bound state is described by the quiver gauge theory on lighter branes \cite{Douglas:1996sw}. The BPS spectra of D-branes in the two limits are generically different. This means that moduli-dependent $\alpha'$-corrections modify the BPS condition for the D-branes. 

For D-branes on a Calabi-Yau three-fold, the BPS spectrum is encoded in the so-called BPS index, which is roughly the Witten index for BPS multiplets. The moduli dependence of the BPS index is called ``wall-crossing phenomena,'' and has recently attracted much attention in physics and mathematics. In particular, a wall-crossing formula which universally characterizes the moduli dependence of BPS indices was proposed in \cite{Denef:2007vg, Kontsevich:2008, Kontsevich:2009xt, Kontsevich:2010, Manschot:2010qz}. Since wall-crossings are highly non-perturbative phenomena in string theory, the discovery of the wall-crossing formula is remarkable.

On the other hand, there are several works \cite{Chuang:2008aw, Chuang:2009pd, Aganagic:2010qr} pointing out that the wall-crossing phenomena of D6-(D4)-D2-D0 states on a Calabi-Yau three-fold are understood in terms of the Seiberg duality \cite{Seiberg:1994pq} of the quiver gauge theory on the D-branes. Here the K\"ahler moduli of the Calabi-Yau geometry are encoded in the FI parameters of the gauge theory \cite{Douglas:1996sw}. Therefore, the moduli dependence of the BPS index is translated into the FI parameter dependence of the Witten index of the quiver gauge theory. An advantage of this approach is that we can use the traditional quantum field theories to study the stringy wall-crossing phenomena of D-branes.

In this paper, we study the wall-crossing phenomena of D4-D2-D0 states on the conifold and orbifold $\mathbb{C}^2/\mathbb{Z}_2$ from the viewpoint of the quiver gauge theory on the D-branes. We put a non-compact D4-brane on a toric divisor of the conifold or orbifold, and consider D2-D0 states bound to it. The quiver gauge theory on the D4-D2-D0 states has recently been identified in \cite{Nishinaka:2013}, but its FI parameter dependence has not been studied yet. Since so far the relation between the Seiberg duality and wall-crossings is limited to D6-(D4)-D2-D0 states, it is worth studying its generalization to the D4-D2-D0 states. Note that the absence of the D6-brane leads to essentially different wall-crossing phenomena. In particular, the D4-D2-D0 states are sensitive to the flop transition unlike the D6-D2-D0 states; the topology of the D4-brane is changed by the flop transition. We particularly study how the flop transition changes the Seiberg duality cascade in the quiver quantum mechanics. Our result turns out to be consistent with the wall-crossing formula for BPS indices.

A further motivation for this study is that the role of the ``anti-quark'' in the D4-D2-D0 quiver was not clear in \cite{Nishinaka:2013}. It was shown in \cite{Nishinaka:2013} that the quiver theory for D4-D2-D0 states includes an ``anti-quark'' $J$, which does not exist in the D6-D2-D0 case.
However, with the values of the FI parameters in \cite{Nishinaka:2013}, the anti-quark $J$ has a vanishing vev on supersymmetric vacua. Therefore, in order to see the physical role of $J$, we have to vary the FI parameters and study the wall-crossing phenomena. We will show that the anti-quark $J$ plays an important role in the Seiberg duality.

We also discuss an interpretation of the Seiberg duality in the associated brane tiling system. Although such an interpretation is well-known for the D2-D0 quivers, the presence of an additional D4-brane needs further study. We claim that the D4-node is always located at an intersection of two special NS5-branes. This leads us to a consistent duality transformation for the D4-D2-D0 states on the orbifold $\mathbb{C}^2/\mathbb{Z}_2$.

The rest of this paper is organized as follows. In section \ref{sec:WC}, we give a brief review of earlier studies on the wall-crossing phenomena of D4-D2-D0 states. In section \ref{sec:D4}, we describe the quiver quantum mechanics associated with the D4-D2-D0 states on the conifold and orbifold singularities. In section \ref{sec:relation}, we discuss the relation between the K\"ahler moduli of the Calabi-Yau three-fold and the FI parameters of the quiver quantum mechanics. In section \ref{sec:duality}, we show how the moduli dependence of the Witten index of the quiver theory is related to the Seiberg duality. In section \ref{sec:dimer}, we give an interpretation of the Seiberg duality with a D4-node in the brane tiling. In section \ref{sec:orbifold}, we apply the prescription given in section \ref{sec:dimer} to the D4-D2-D0 states on $\mathbb{C}^2/\mathbb{Z}_2$. In section \ref{sec:discussion}, we give a concluding remark.

\section{The wall-crossing phenomena of D4-D2-D0 states}
\label{sec:WC}

In this paper, we consider a single D4-brane on a toric Calabi-Yau three-fold $Y$, and BPS D2-D0 states bound to it. We particularly focus on $Y$ which has a single compact two-cycle and no compact four-cycle. Such $Y$ is written as a total space of $\mathcal{O}(-N)\oplus \mathcal{O}(N-2)\to \mathbb{P}^1$ for $N=0,1$ or $2$. If $N=1$ then $Y$ is the resolved conifold, while if $N=0,2$ then $Y$ is the trivial line bundle over the $A_1$-ALE space. We assume that the D4-brane is put on a toric divisor of $Y$, the D2-branes are wrapped on the compact two-cycle of $Y$, and the D0-branes are point-like in $Y$. The spectrum of the BPS D4-D2-D0 states depends on the K\"ahler moduli of $Y$. In this section, we briefly review the works \cite{Nishinaka:2010qk, Nishinaka:2011nn} on the moduli dependence of the D4-D2-D0 spectrum on $Y$.\footnote{See also \cite{Nishinaka:2010fh, Nishinaka:2011sv, Nishinaka:2011pd, Nishinaka:2011is} for more on the related wall-crossing phenomena of D4-D2-D0 states. For the D6-D2-D0 wall-crossings on $Y$, see \cite{Jafferis:2008uf,Chuang:2010wx}. }

When we dimensionally reduce $Y$, we obtain a $d=4,\mathcal{N}=2$ supersymmetric $U(1)$ gauge theory in which the D4-D2-D0 bound state is regarded as a BPS particle. The Ramond-Ramond charge of the D-branes is then identified with the electro-magnetic charge $\gamma$ of the BPS particle. In $d=4,\mathcal{N}=2$ theories, the so-called ``BPS index'' captures the spectrum of stable BPS bound states.
 For the basic property of the BPS index, see appendix \ref{app:index}.
Since the spectrum of the D4-D2-D0 states depends on the K\"ahler moduli of $Y$, so does the BPS index $\Omega(\gamma)$. In the $d=4,\mathcal{N}=2$ theory, the K\"ahler moduli of $Y$ is interpreted as the vector multiplet moduli. Since the BPS index is integer-valued, it can only change discontinuously. A discontinuous change of the BPS index under a variation of the vector multiplet moduli is called a ``wall-crossing phenomenon.''

Let us denote by $D$ the toric divisor wrapped by the D4-brane. We consider $N_2$ D2-branes and $N_0$ D0-branes bound to the D4-brane. When $N_2<0$ or $N_0<0$, we regard them as anti D2 or anti D0-branes, respectively. We express the D-brane charge in terms of a differential form on $Y$:
\begin{eqnarray}
\gamma = -\mathcal{D} + N_2\beta - N_0 dV,
\label{eq:charge}
\end{eqnarray}
where $\mathcal{D}$ is the two-form dual to $D$, $\beta$ is the four-form dual to the base $\mathbb{P}^1$, and $dV$ is the volume-form. The BPS index $\Omega(\gamma)$ essentially counts the number of stable BPS one-particle states carrying charge $\gamma$. For later use, we define the BPS partition function $\mathcal{Z}$ by
\begin{eqnarray}
\mathcal{Z} = \sum_{N_0,N_2\in \mathbb{Z}}\Omega(\mathcal{D}+N_2\beta-N_0 dV)q^{N_0}Q^{N_2}P,
\label{eq:partition}
\end{eqnarray}
where $q,Q$ and $P$ are fugacities associated with the D0-, D2- and D4-charges. Since our D4-charge is always one, we usually set $P=1$.

Since BPS states are usually stable against a small variation of parameters, the BPS index $\Omega(\gamma)$ is piecewise constant in the moduli space. However, in a special subspace of the moduli space, the index $\Omega(\gamma)$ can in fact {\it jump.} This is called the ``wall-crossing phenomenon.'' A wall-crossing is possible only if there is a decay channel $\gamma\to \gamma_1+\gamma_2$ of a BPS state with charge $\gamma$. Note that the charge conservation implies $Z(\gamma) = Z(\gamma_1) + Z(\gamma_2)$, which particularly requires $|Z(\gamma)|\leq |Z(\gamma_1)| + |Z(\gamma_2)|$. On the other hand, the BPS bound and energy-momentum conservation imply $|Z(\gamma)|\geq |Z(\gamma_1)| + |Z(\gamma_2)|$. Hence, the decay channel is possible only if
 \begin{eqnarray}
\arg Z(\gamma_1) = \arg Z(\gamma_2).
\label{eq:wall3}
\end{eqnarray}
Note that, even if this condition is satisfied, the decay is a marginal decay $\gamma\leftrightarrow \gamma_1+\gamma_2$.
 Since the central charge implicitly depends on the vacuum moduli, we can solve \eqref{eq:wall3} for the moduli parameters. The solution space is a real codimension one subspace in the moduli space, which is called a ``wall of marginal stability.'' We have a wall for each possible decay channel. The moduli space is divided into chambers surrounded by the walls. The BPS index $\Omega(\gamma)$ is constant in each chamber, but can jump when the moduli cross one of the walls.

To identify the walls of marginal stability in our setup, let us first evaluate the central charges of the D4-D2-D0 states.  The central charge $Z(\gamma)$ is a linear function of $\gamma$. Since \eqref{eq:wall3} is independent of the normalization of $Z$, we normalize it so that a single D0-brane has
\begin{eqnarray}
Z(-dV) = 1.
\end{eqnarray}
The central charge of the D2-brane depends on the complexified K\"ahler moduli of $Y$. We take the coordinate $z$ of the moduli space so that the central charge of a single D2-brane is given by
\begin{eqnarray}
Z(\beta) = z.
\end{eqnarray}
This implies that, in the large radius limit of $\mathbb{P}^1$, ${\rm Re}\,z$ and $|{\rm Im}\,z|$ coincide with the B-field and area of the $\mathbb{P}^1$, respectively. The point $z=0$ is a singularity in the moduli space, at which the D2-branes become massless.
The central charge of the D4-brane is divergent because it is wrapped on the non-compact divisor $D$. We therefore regularize it as
\begin{eqnarray}
Z(-\mathcal{D}) = -\frac{1}{2}\Lambda^2 e^{2i\varphi},
\end{eqnarray}
where $\Lambda$ and $\varphi$ are real parameters.
 In the final expression, we should take the limit $\Lambda\to \infty$. This type of regularization was first given in \cite{Jafferis:2008uf}. The phase $\varphi$ expresses the ``ratio'' of the B-field and volume of the divisor $D$, and specifies which half of the supersymmetry the D4-brane breaks. In this paper, we fix $\varphi$ so that $\frac{\pi}{4}<\varphi<\frac{\pi}{2}$ as in \cite{Nishinaka:2010qk, Nishinaka:2011nn}.

Now, let us solve \eqref{eq:wall3} to identify the locations of the walls of marginal stability. For our charge \eqref{eq:charge}, any wall-crossing is associated with a decay channel $\gamma\to\gamma_1+\gamma_2$ for
\begin{eqnarray}
\gamma_1 = m\beta -ndV,\qquad \gamma_2 = \gamma-\gamma_1,
\label{eq:channel}
\end{eqnarray}
with some $m,n\in\mathbb{Z}$.
The reason for this is that, since $Y$ is a non-compact three-fold without compact four-cycles, any pair creation of D6-$\overline{\rm D6}$ or D4-$\overline{\rm D4}$ is forbidden by the energy conservation \cite{Jafferis:2008uf,Nishinaka:2010qk}. We denote by $W_n^m$ the wall associated with the decay channel \eqref{eq:channel}. Note here that the channel \eqref{eq:channel} is forbidden if there is no stable D2-D0 state with charge $\gamma_1$. The equation \eqref{eq:wall3} implies that the wall $W_n^m$ is located in the subspace
\begin{eqnarray}
{\rm arg}(-m z - n) = 2\varphi  \quad \text{mod}\quad 2\pi
\end{eqnarray}
in the moduli space.
When the moduli cross one of the walls $W^m_n$, the BPS index could be changed.
 It was proposed in \cite{Denef:2007vg, Kontsevich:2008, Kontsevich:2009xt, Kontsevich:2010, Manschot:2010qz} that the changes of the BPS indices are characterized by the so-called wall-crossing formula. In our setup, the formula implies that the BPS partition function $\mathcal{Z}$ changes as
\begin{eqnarray}
\mathcal{Z} \to \mathcal{Z}\times \prod_{k=1}^\infty(1-q^{kn}Q^{km})^{\pm k\langle \gamma_1,\gamma_2\rangle \Omega(k\gamma_1)}
\label{eq:WC_formula}
\end{eqnarray}
 when the moduli cross the wall $W^m_n$. The sign of the exponent depends on from which side the moduli cross the wall. This particularly implies that the BPS index is invariant if the charge intersection product $\langle \gamma_1,\gamma_2\rangle$ vanishes. In particular, the walls $W^0_n$ give rise to no change in the BPS index $\Omega(\gamma)$.

\subsection*{Resolved conifold}

When $N=1$, the Calabi-Yau three-fold $Y$ is the resolved conifold. We put a D4-brane on a divisor $D=(\mathcal{O}(-1)\to \mathbb{P}^1)$, and consider BPS D2-D0 states bound to it. The non-vanishing BPS index for $\gamma_1 = m\beta-ndV$ is read off as
\begin{eqnarray}
\Omega(\gamma_1) = 1 \quad \text{if} \quad m=\pm1,0,
\label{eq:D2D0-conifold}
\end{eqnarray}
from the Gopakumar-Vafa invariants of the conifold. Unless this is non-vanishing, the corresponding wall $W^m_n$ does not change the partition function through \eqref{eq:WC_formula}.
Therefore only the walls $W^{\pm1}_n$ need to be considered.\footnote{As already mentioned, the walls $W^0_n$ give rise to no wall-crossing, because of the vanishing charge intersection product.} 
Note here that $\Omega(\gamma_1)$ is {\it exactly constant} in the whole moduli space. The reason is again that $Y$ is a non-compact three-fold without compact four-cycles. Since any D4-$\overline{\text{D4}}$ or D6-$\overline{\text{D6}}$ pair creation is forbidden, the only possible decay channels are separations of D2-D0 fragments. However, a D2-D0 separation does not change $\Omega(\gamma_1)$ because the charge intersection product between D2-D0 states always vanishes.

\begin{figure}
\begin{center}
\includegraphics[width=11cm]{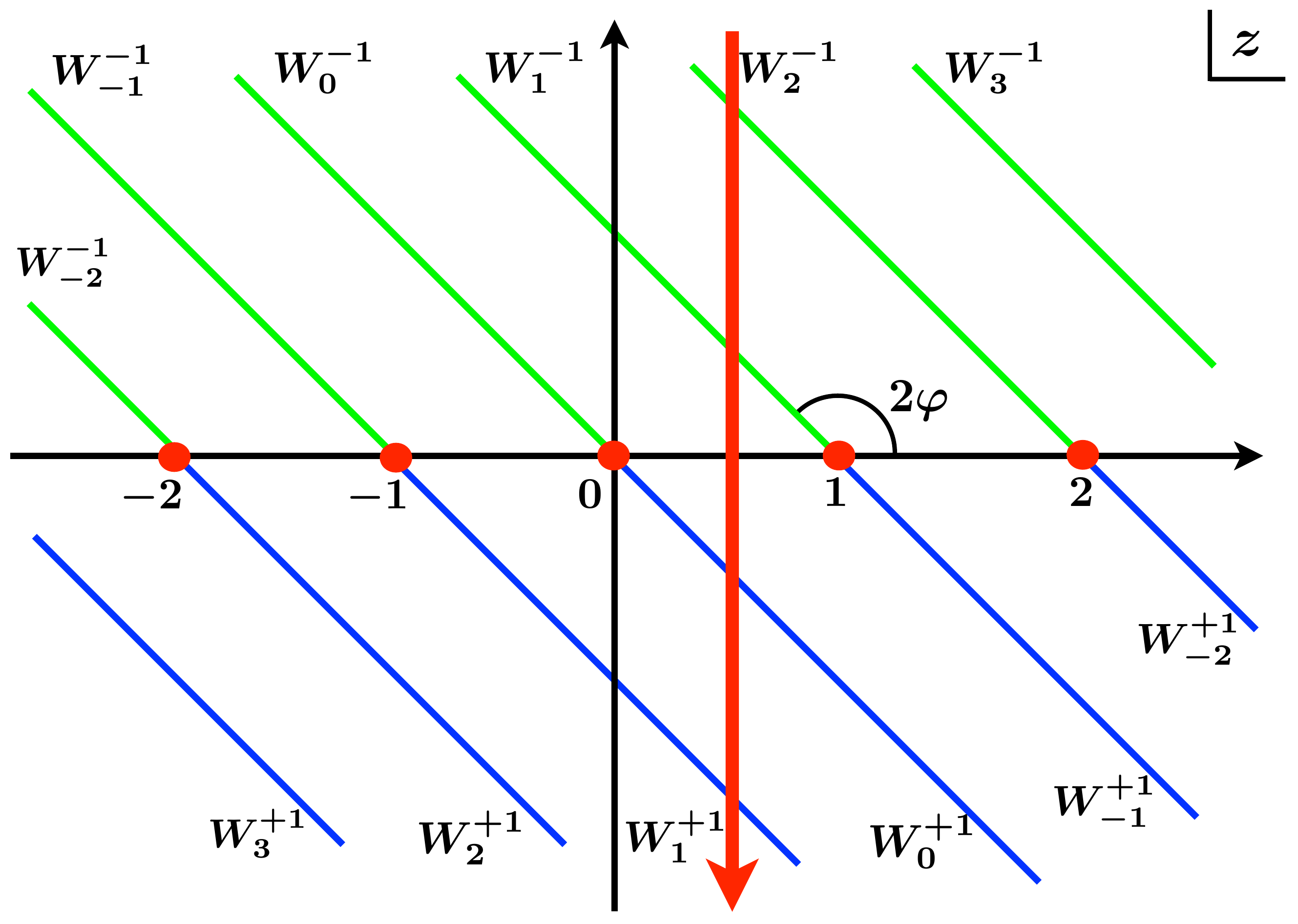}
\caption{The walls of marginal stability in the $z$-plane. Each red dot is a singularity at which some D2-D0 state becomes massless. Throughout this paper, we avoid crossing the singularities.}
\label{fig:wall}
\end{center}
\end{figure}

The locations of the walls $W^{\pm1}_n$ are shown in figure \ref{fig:wall}.
Here, the moduli region ${\rm Im}\,z>0$ and ${\rm Im}\,z<0$ are connected by the flop transition.
The moduli space is now divided into an infinite number of chambers. For $n\in \mathbb{Z}$, we denote by $C_n$ the chamber between $W^{+1}_{n-1}$ and $W^{+1}_{n}$, or equivalently between $W^{-1}_{-n}$ and $W^{-1}_{-n+1}$. The condition for $z$ to be in $C_n$ is written as
\begin{eqnarray}
{\rm Im}\left((z+n)e^{-2i\varphi}\right)<0,\qquad {\rm Im}\left((1-n-z)e^{-2i\varphi}\right)<0.
\label{eq:Ck}
\end{eqnarray}
Note that the flop transition itself does not change the partition function $\mathcal{Z}$.
The partition function $\mathcal{Z}$ in each chamber was evaluated in \cite{Nishinaka:2010qk}, by using the wall-crossing formula \eqref{eq:WC_formula}. When we move the moduli parameter $z$ along the red arrow in figure \ref{fig:wall}, the partition function is written as
\begin{eqnarray}
\mathcal{Z} = \prod_{k=1}^\infty\frac{1}{1-q^k}\prod_{\ell=0}^\infty(1-q^\ell Q)\prod_{m=1}^{|n|}(1-q^{m}Q^{-1})
\label{eq:formula}
\end{eqnarray}
in the chamber $C_{n<0}$, and written as
\begin{eqnarray}
\mathcal{Z} = \prod_{k=1}^\infty\frac{1}{1-q^k}\prod_{\ell=n}^\infty(1-q^\ell Q)
\label{eq:formula2}
\end{eqnarray}
 in the chamber $C_{n\geq 0}$. Note that we already set $P=1$ both in \eqref{eq:formula2} and \eqref{eq:formula}. In section \ref{sec:duality}, we rederive these results from the quiver quantum mechanics on the D-branes, without using the wall-crossing formula \eqref{eq:WC_formula}.

\subsection*{Resolved $\mathbb{C}^2/\mathbb{Z}_2$}

When $N=0$ or $2$, the Calabi-Yau three-fold $Y$ is a trivial line bundle over the $A_1$-ALE space. We put a D4-brane on the ALE space, i.e. $D = (\mathcal{O}(-2)\to \mathbb{P}^1)$. The Gopakumar-Vafa invariants tell us that the non-vanishing BPS index for charge $\gamma_1$ is
\begin{eqnarray}
\Omega(\gamma_1) = 
\left\{
\begin{array}{l}
-1 \quad \text{if}\quad m=\pm1\\
+1 \quad \text{if}\quad m=0\\
\end{array}
\right..
\label{eq:D2D0-orbifold}
\end{eqnarray}
As in the conifold case, $\Omega(\gamma_1)$ is exactly constant in the moduli space.

The K\"ahler modulus $z$ is associated with the base $\mathbb{P}^1$. Since $Y$ does not admit the flop transition, we restrict $z$ so that ${\rm Im}\,z\geq 0$. The small radius limit of the base $\mathbb{P}^1$ corresponds to $0<{\rm Im}\,z \ll 1$. The BPS index \eqref{eq:D2D0-orbifold} implies that the charge $\gamma_1$ does not give any wall-crossing unless $m=\pm1$. However, the wall-crossing formula \eqref{eq:WC_formula} now tells us that even the walls $W^{\pm1}_n$ give rise to no wall-crossings. The reason for this is that the charge intersection product between $\gamma_1$ and $\gamma_2$ always vanishes:
\begin{eqnarray}
\langle \gamma_1, \gamma_2\rangle = 0.
\end{eqnarray}
This is a consequence of the fact that the base $\mathbb{P}^1$ is not rigid in $Y=(\mathcal{O}(-2)\oplus\mathcal{O}(0)\to \mathbb{P}^1)$. Therefore, {\it there are no wall-crossing phenomena in this setup, and the partition function $\mathcal{Z}$ is exactly constant in the $z$-plane.}

In the large radii limit, the partition function $\mathcal{Z}$ is identified with the partition function of the field theory on the D4-brane. The field theory is the Vafa-Witten theory on the $A_1$-ALE space, whose partition function is written as \cite{Vafa:1994tf, Nakajima:1994,  Aganagic:2004js, Szabo:2009vw}
\begin{eqnarray}
\mathcal{Z} = \prod_{k=1}^\infty\left(\frac{1}{1-q^k}\right)^2\sum_{\ell\in\mathbb{Z}}q^{\ell^2}Q^{\ell} = \frac{q^{1/12}}{\eta(q)}\chi^{\widehat{su}(2)_1}(q,Q),
\label{eq:affine}
\end{eqnarray}
where we set $P=1$ as before.
Note that this is proportional to the character of $\widehat{su}(2)_1$.\footnote{To be precise, this is the character for the trivial level-one weight of affine $SU(2)$ algebra. The character for the non-trivial level-one weight is obtained by turning on the non-trivial holonomy of the gauge field at infinity on the D4-brane.}
Since there are no wall-crossings, the BPS partition function $\mathcal{Z}$ has the same expression at any point in the $z$-plane. In section \ref{sec:orbifold}, we rederive this result from the quiver quantum mechanics on the D-branes, without using the wall-crossing formula \eqref{eq:WC_formula}.

\section{Quiver for D4-D2-D0 at singularity}
\label{sec:D4}

Near the singular point $z=0$ in the moduli space, the D2-branes are realized as fractional D0-branes. The low-energy dynamics of the D-branes is then described by a quiver gauge theory \cite{Douglas:1996sw}. In our setup, the quiver theory lives on the one-dimensional world-line of the BPS particle. Since the BPS particle breaks half the eight supersymmetries of the target space, the quiver theory is a $d=1,\mathcal{N}=4$ quiver quantum mechanics. In the $d=4,\mathcal{N}=1$ language, the theory is characterized by the quiver diagram $Q$ and superpotential $W$. The quiver $Q$ and superpotential $W$ for our D4-D2-D0 states were identified in \cite{Nishinaka:2013}. In this section, we simply quote the result.

\subsection{Conifold}

\begin{figure}
\begin{center}
\includegraphics[width=4cm]{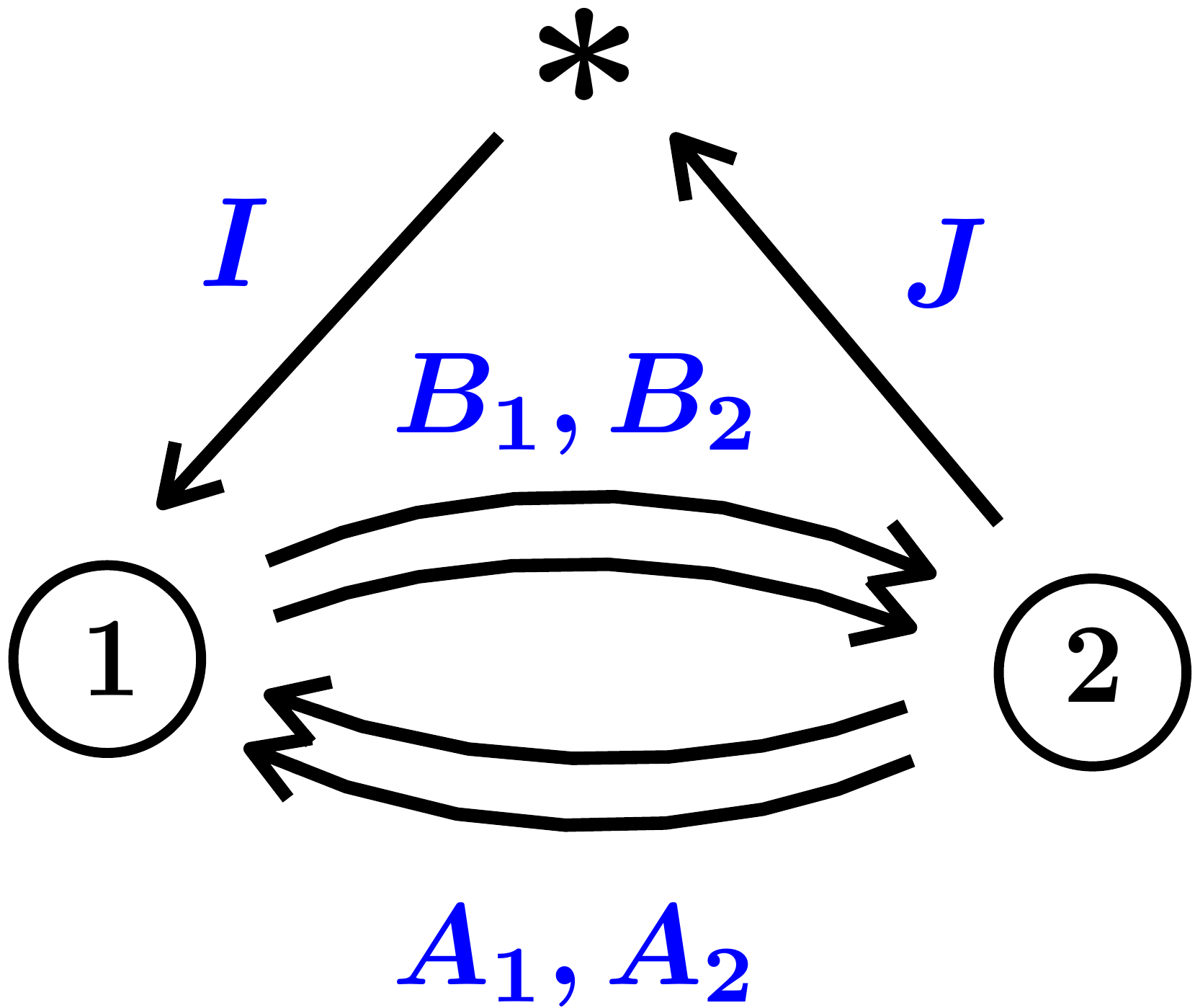} 
\caption{The quiver diagram for the D4-D2-D0 states on the conifold.}%
\label{fig:quiver-conifold}
\end{center}
\end{figure}
When the D4-D2-D0 states are on the conifold singularity, the quiver diagram $Q$ can be depicted as in figure \ref{fig:quiver-conifold}. We have three quiver nodes denoted by $1,2$ and $*$. A D4-D2-D0 state is expressed as a {\it representation} of this quiver $Q$. Given a representation $R$ of the quiver $Q$, each quiver node is associated with a vector space over $\mathbb{C}$, which we denote by $V_1,V_2$ and $V_*$ respectively. The dimensions of the vector spaces are related to the D-brane charge $\gamma = -\mathcal{D}+N_2\beta -N_0 dV$ as
\begin{eqnarray}
{\rm dim}\,V_1 = N_2+N_0,\qquad {\rm dim}\,V_2 = N_0,\qquad {\rm dim}\,V_* = 1.
\label{eq:rank}
\end{eqnarray}
Therefore, the node $*$ is associated with the D4-charge while the other nodes are associated with the D2-D0 charge.
Let us consider the basis representation $R^{(k)}$ which has $\dim V_k=1$ and $\dim V_{\ell}=0$ for $\ell\neq k$. We denote by $\gamma^{(k)}$ the D-brane charge associated with the representation. Then the relation \eqref{eq:rank} is equivalent to
\begin{eqnarray}
\gamma^{(1)} = \beta,\qquad \gamma^{(2)} = -\beta-dV,\qquad \gamma^{(*)} = -\mathcal{D}.
\label{eq:unit0}
\end{eqnarray}
The charge $\gamma^{(k)}$ is the unit charge associated with the quiver node $k$. Each quiver node $k$ is also associated with a gauge group $U({\rm dim}\,V_k)$ and FI parameter $\theta_k$.

The theory has six chiral multiplets $A_{1},A_{2}, B_{1}, B_{2}, I$ and $J$ with the superpotential
\begin{eqnarray}
W= {\rm tr}(A_1B_1A_2B_2) - {\rm tr}(A_1B_2A_2B_1) + JB_2I.
\label{eq:superpot}
\end{eqnarray}
Note that the first two terms of \eqref{eq:superpot} are the well-known Klebanov-Witten type potential \cite{Klebanov:1998hh, Klebanov:2000hb}. This is reasonable because, by taking the T-duality along three spacial directions transverse to the conifold, the D2-D0 states become D5-D3 states studied in \cite{Klebanov:1998hh, Klebanov:2000hb}. On the other hand, the third term in \eqref{eq:superpot} is induced by the presence of the D4-brane.

Now, a BPS configuration of the D4-D2-D0 state corresponds to a supersymmetric vacuum of the quiver quantum mechanics, i.e. a solution to the F- and D-term constraints. The F-term constraints are given by 
\begin{eqnarray}
\frac{\partial W}{\partial X} = 0 
\end{eqnarray}
for all the chiral multiplet $X$, while the D-term constraints are given by
\begin{eqnarray}
\sum_{X\in S_k}X^\dagger X - \sum_{X\in T_k} XX^\dagger = \theta_k{\bf 1}
\label{eq:D-term3}
\end{eqnarray}
for $k=1,2,*$. Here $S_k$ and $T_k$ are the set of arrows from and to the node $k$, respectively.
Note that $\theta_{1,2,*}$ are not all independent; summing up the traces of \eqref{eq:D-term3} for all $k$ leads to
\begin{eqnarray}
\theta_1\, {\rm dim}V_1 + \theta_2\, {\rm dim}V_2 + \theta_*\, {\rm dim}V_* = 0.
\label{eq:theta_zero}
\end{eqnarray}
 This implies that, for a given representation $R$, we have two independent FI parameters.

The BPS index $\Omega(\gamma)$ of the D-brane is now identified with the Witten index of the quiver quantum mechanics. Therefore, the BPS partition function \eqref{eq:partition} corresponds to the generating function of the Witten index. For the FI parameters such that 
\begin{eqnarray}
\theta_1 < 0,\qquad \theta_2<0,\qquad \theta_*\geq 0,
\label{eq:identification}
\end{eqnarray}
the generating function of the Witten index was evaluated in \cite{Nishinaka:2013} as
\begin{eqnarray}
\mathcal{Z} = w\prod_{n=1}^\infty \frac{1}{1-(xy)^n}\prod_{m=0}^\infty (1-(xy)^nx).
\label{eq:partition0}
\end{eqnarray}
Here $x,y$ and $w$ are fugacities associated with the unit charges $\gamma^{(1)},\gamma^{(2)}$ and $\gamma^{(*)}$, respectively. The relation \eqref{eq:unit0} implies that they are related to $q,Q$ and $P$ by
\begin{eqnarray}
x = Q,\qquad y = qQ^{-1},\qquad w= P=1.
\label{eq:fugacity1}
\end{eqnarray}
Since the FI parameters are related to the K\"ahler moduli of the background Calabi-Yau three-fold \cite{Douglas:1996sw}, varying the FI parameters is expected to give rise to wall-crossing phenomena. In \cite{Nishinaka:2013}, it was assumed that the FI parameters satisfying \eqref{eq:identification} correspond to the chamber $C_0$. In fact, by substituting \eqref{eq:fugacity1}, the generating function \eqref{eq:partition0} perfectly agrees with $\mathcal{Z}$ for $k=0$ in \eqref{eq:formula2}. We will verify this assumption in the next section.

\subsection{Orbifold $\mathbb{C}^2/\mathbb{Z}_2$}
\label{subsec:orbifold}

\begin{figure}
\begin{center}
\includegraphics[width=6.5cm]{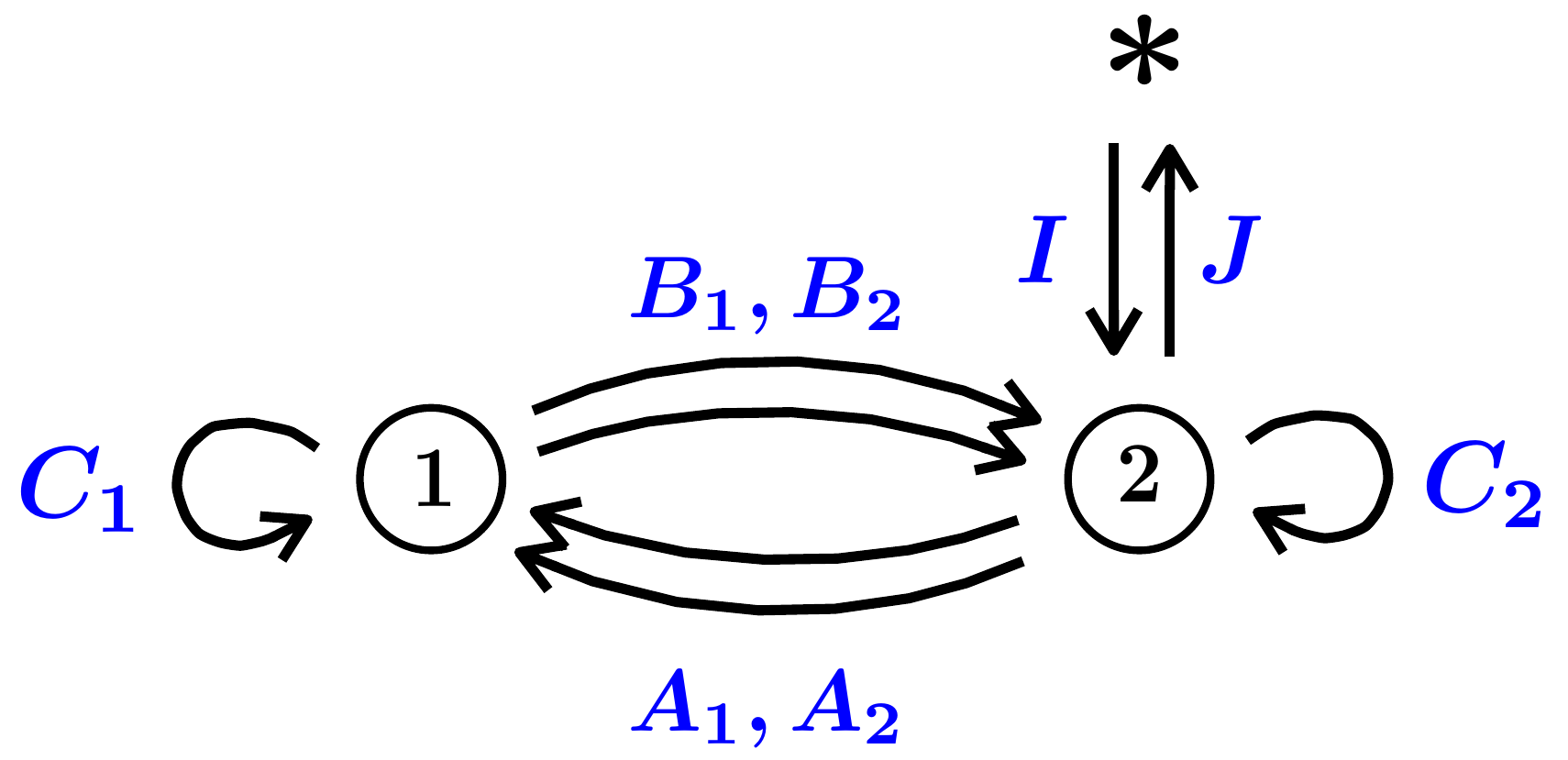}
\caption{The quiver diagram for the D4-D2-D0 states on the orbifold $\mathbb{C}^2/\mathbb{Z}_2$.}
\label{fig:quiver-orbifold}
\end{center}
\end{figure}

The quiver theory for the D4-D2-D0 states on the orbifold $\mathbb{C}^2/\mathbb{Z}_2$ was also studied in \cite{Nishinaka:2013}. Its quiver diagram is shown in figure \ref{fig:quiver-orbifold}. The theory involves three quiver nodes as well as eight chiral multiplets $A_1,A_2,B_1,B_2,C_1,C_2,I$ and $J$. The identification between the charge $\gamma$ and the rank of the gauge group is the same as in \eqref{eq:rank} and \eqref{eq:unit0}. The superpotential is given by
\begin{eqnarray}
W = {\rm tr}\left(C_1(A_1B_1 - A_2B_2)\right) - {\rm tr}\left(C_2(B_1A_1 - B_2A_2)\right) + JC_2I.
\label{eq:superpot-orbifold}
\end{eqnarray}
In particular, the third term $JC_2I$ is induced by the presence of the D4-brane.

There are three FI parameters $\theta_{1,2,*}$ constrained by \eqref{eq:theta_zero}. For the FI parameters satisfying \eqref{eq:identification}, the generating function of the Witten index was evaluated in \cite{Nishinaka:2013} as
\begin{eqnarray}
\mathcal{Z} = w\prod_{n=1}^\infty\frac{1}{1-(xy)^n}\sum_{m\in\mathbb{Z}}(xy)^{n^2}x^{n}.
\label{eq:partition2}
\end{eqnarray}
The fugacities $x,y$ and $w$ are again associated with the unit charge $\gamma^{(1)}, \gamma^{(2)}$ and $\gamma^{(*)}$, respectively. The relation between the fugacities $x,y,w$ and $q,Q,P$ is given by \eqref{eq:fugacity1}. Note that, when substituting \eqref{eq:fugacity1}, the generating function \eqref{eq:partition2} coincides with \eqref{eq:affine}. In \cite{Nishinaka:2013}, the FI parameters satisfying \eqref{eq:identification} was assumed to correspond to the small radius limit of $\mathbb{P}^1$. We will verify this assumption in the next section.

\section{Relation between K\"ahler and FI parameters}
\label{sec:relation}

Once we fix the unit charges associated with the quiver nodes, the quiver description of the D-brane bound states is reliable only in a subspace of the moduli space. For a general value of the moduli parameter, there could exist a stable D-brane bound state which is not expressed as a single quiver representation. This is similar to the fact that $(p+1)$-dimensional field theory description of D$_p$-branes is valid only near the large radii limit of compact cycles.

A general bound state of D-branes is expressed as a {\it complex} of quiver representations. In other words, it is an object of the derived category of quiver representations \cite{Kontsevich:1994, Douglas:2000ah, Douglas:2000qw, Douglas:2000gi}. For example, let us consider two quiver representations $\mathcal{S}$ and $\mathcal{S}'$ associated with some D4-D2-D0 states. Then the bound state of $\mathcal{S}$ and $\overline{\mathcal{S}'}$ is expressed as
\begin{eqnarray}
\cdots \longrightarrow 0 \longrightarrow \mathcal{S}'\xrightarrow{\,\;f\;\,} \mathcal{S} \longrightarrow 0 \longrightarrow \cdots,
\end{eqnarray}
where $\mathcal{S}$ sits at the zero-th position. The morphism $f$ is associated with a tachyonic open string between $\mathcal{S}$ and $\overline{\mathcal{S}'}$. In general, any complex with finite length expresses some bound state of D-branes. We say that a complex is {\it stable} if the corresponding bound state is stable.

If every stable complex is quasi-isomorphic to a complex
\begin{eqnarray}
\cdots \longrightarrow 0 \longrightarrow \mathcal{S}'' \longrightarrow 0 \longrightarrow \cdots
\label{eq:complex}
\end{eqnarray}
for some quiver representation $\mathcal{S}''$, then we do not need to consider the derived category; it is sufficient to consider quiver representations $\mathcal{S}''$.
A sufficient condition for this was given in \cite{Aspinwall:2004mb}. Suppose that we have a quiver $Q$ with $n$ nodes. Then, for $k=1,\cdots,n$, the basis representation $R^{(k)}$ carries the unit charge $\gamma^{(k)}$ associated with the $k$-th quiver node. It was shown in \cite{Aspinwall:2004mb} that, if all the representations $R^{(k)}$ are stable and all the central charges $Z(\gamma^{(k)})$ lie in a {\it convex cone} in the complex plane, then every stable complex is quasi-isomorphic to \eqref{eq:complex} for some $\mathcal{S}''$. This is physically interpreted to mean that the tachyon condensation reduces any bound state of D-branes to some single $\mathcal{S}''$. Moreover it was shown that, in this case, the stable complexes are in one-to-one correspondence with the solutions to the F- and D-term constraints.\footnote{To be precise, \cite{Aspinwall:2004mb} showed that the $\Pi$-stability reduces to King's $\theta$-stability. The work \cite{King:1994} implies that the $\theta$-stable representations are in one-to-one correspondence with the D-flat configurations. The F-flatness is already realized as the ``relation'' of the quiver representation.}

\begin{figure}
\begin{center}
\includegraphics[width=10cm]{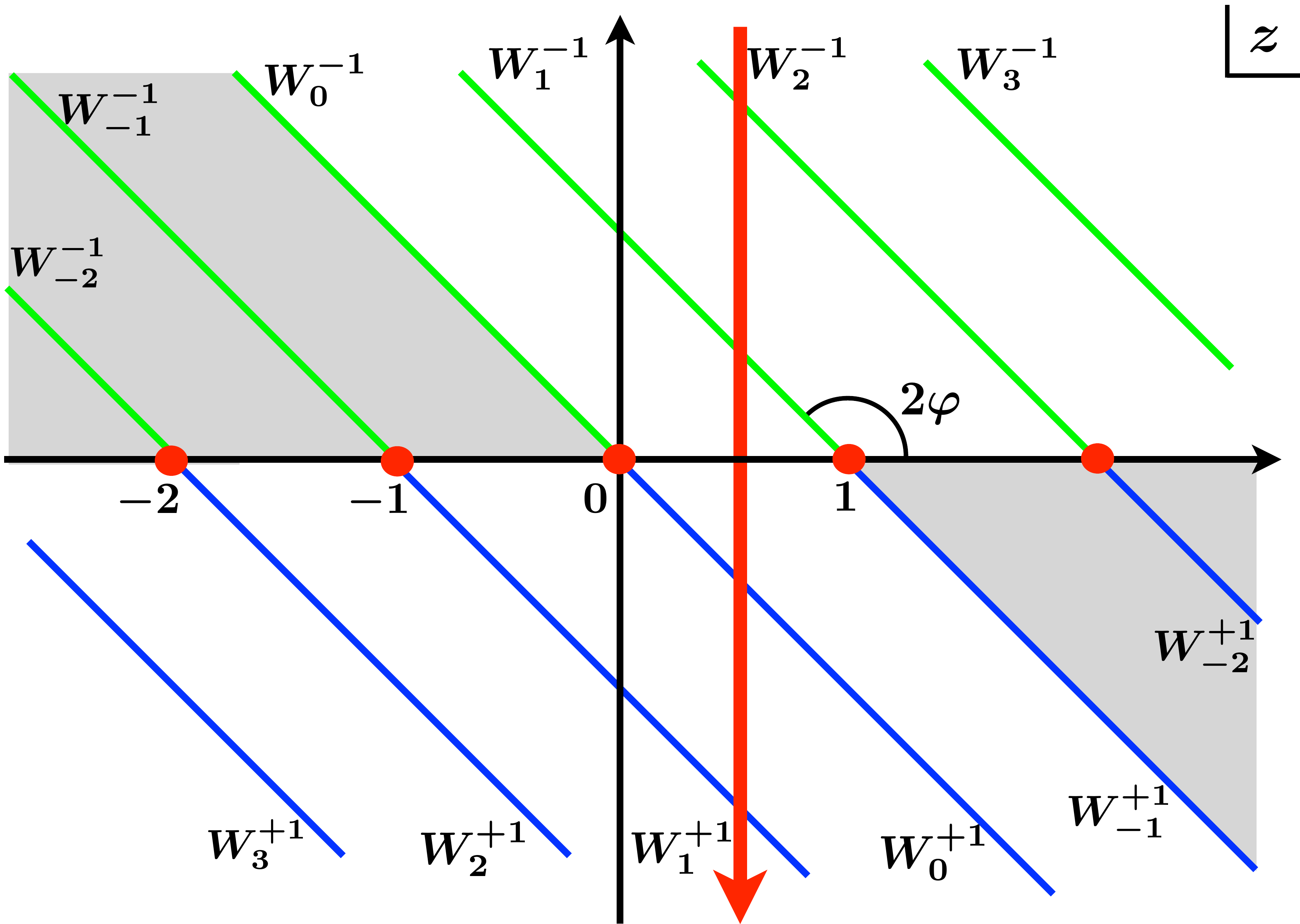}
\caption{In the {\it un-shaded} region, the quiver quantum mechanics in section 3 describe the whole D4-D2-D0 states. In particular, along the red arrow, the quiver quantum mechanics description is reliable.}
\label{fig:region1}
\end{center}
\end{figure}

In our case, there are three quiver nodes $1,2$ and $*$. The corresponding unit charges are given by \eqref{eq:unit0}. Since $\gamma^{(*)}$ is the charge of the non-compact D4-brane, $R^{(*)}$ is stable in the whole moduli space.\footnote{Note that, if the D4-brane is wrapped on a compact four-cycle of a compact Calabi-Yau three-fold, then $R^{(*)}$ might decay into a D6-$\overline{\text{D6}}$ pair.} On the other hand, it follows from \eqref{eq:D2D0-conifold} and \eqref{eq:D2D0-orbifold} that the BPS indices of $\gamma^{(1)}$ and $\gamma^{(2)}$ are both non-zero at any point of the moduli space. Therefore, $R^{(1)}$ and $R^{(2)}$ are also stable. The central charges for $\gamma^{*,1,2}$ are given by
\begin{eqnarray}
Z(\gamma^{(*)}) = -\frac{1}{2}\Lambda^2e^{2i\varphi},\qquad Z(\gamma^{(1)}) = z,\qquad Z(\gamma^{(2)}) = -z+1.
\label{eq:Ccharge}
\end{eqnarray}
In order for $Z(\gamma^{(*,1,2)})$ to lie in a convex cone in the complex plane, we have to restrict the value of $z$ in the {\it un-shaded} region in figure \ref{fig:region1}. In particular, when the moduli move along the red arrow, we can trust the quiver quantum mechanics description of the D4-D2-D0 states. Therefore, we focus on the region $0<{\rm Re}\,z<1$ in the rest of this paper.

When the quiver quantum mechanics description is reliable, the K\"ahler moduli of $Y$ are translated into the FI parameters. The explicit relation is as follows \cite{Douglas:2000ah, Douglas:2000qw, Fiol:2000pd, Aspinwall:2004mb}. Suppose that we are interested in the D4-D2-D0 state associated with a quiver representation $R$. We call $R$ the ``reference representation.'' We fix the FI parameters $\theta_k$ so that \eqref{eq:theta_zero} holds on the reference representation $R$. We then define
\begin{eqnarray}
\theta(R') = \theta_1\, {\rm dim}\,V'_1 + \theta_2\,{\rm dim}\, V'_2  + \theta_*\,{\rm dim}\,V'_*
\label{eq:theta-R'}
\end{eqnarray}
for a quiver representation $R'$. Here $V'_{k}$ is the vector space associated with the $k$-th quiver node in the representation $R'$.
 Then, the relation between the FI and K\"ahler parameters are given by
\begin{eqnarray}
\theta(R')  = -{\rm Im}\,\frac{Z(\gamma_{R'})}{Z(\gamma_{R})},
\end{eqnarray}
where $\gamma_{R'}$ is the D-brane charge associated with the representation $R'$. 
Note that this properly satisfies $\theta(R) = 0$. To be more explicit, suppose that $\dim V'_1 = p,\;\dim V'_2 = q$ and $\dim V'_* = 1$ for the representation $R'$. Then the corresponding central charge is written as
\begin{eqnarray}
Z(\gamma_{R'}) = -\frac{1}{2}\Lambda^2e^{2i\varphi}  + (p-q) z + q,
\end{eqnarray}
which implies
\begin{eqnarray}
\theta(R')
=  p\,{\rm Im}\left(\frac{-z}{Z(\gamma_{R})}\right) + q\,{\rm Im}\left(\frac{z-1}{Z(\gamma_{R})}\right) +\, {\rm Im}\left(\frac{\frac{1}{2}\Lambda^2 e^{2i\varphi}}{Z(\gamma_{R})}\right).
\label{eq:theta}
\end{eqnarray}
Here, the charge $\gamma_{R}$ of the reference representation $R$ is given by $\gamma$ in \eqref{eq:charge}. By comparing \eqref{eq:theta} with \eqref{eq:theta-R'}, we obtain the relation between the FI  and K\"ahler parameters. In particular, in the limit of $\Lambda\to \infty$, we obtain
\begin{eqnarray}
\theta_1 = \frac{2}{\Lambda^2}{\rm Im}\left(ze^{-2i\varphi}\right),\quad \theta_2 = \frac{2}{\Lambda^2}{\rm Im}\left((1-z)e^{-2i\varphi}\right),\quad \theta_* = -\frac{2}{\Lambda^2}{\rm Im}\left((N_2z+N_0)e^{-2i\varphi}\right).
\nonumber\\
\label{eq:FI-K}
\end{eqnarray}

Recall that it was assumed in \cite{Nishinaka:2013} that $\theta_1,\theta_2<0$ corresponds to the small radius limit of the compact two-cycle. We can now prove this statement. From \eqref{eq:FI-K}, the condition $\theta_1,\theta_2<0$ is equivalent to
\begin{eqnarray}
 {\rm Im}\left(ze^{-2i\varphi}\right)<0, \qquad {\rm Im}\left((1-z)e^{-2i\varphi}\right)<0.
\label{eq:C0}
\end{eqnarray}
The solution space to these equations exactly coincides with the chamber $C_0$ defined in section \ref{sec:WC}! Since we have the restriction $0<{\rm Re}\,z<1$ on the moduli parameter, the chamber $C_0$ corresponds to the small radius limit of the compact two-cycle, i.e. $|{\rm Im}\,z| \ll 1$.

When the moduli parameter $z$ is outside the chamber $C_0$, the FI parameters break the condition $\theta_1,\theta_2<0$. The result in section \ref{sec:D4} is no longer available due to the wall-crossing phenomena. In the rest of this paper, we discuss how to describe such a moduli region in terms of the quiver quantum mechanics.

\section{Seiberg duality and wall-crossings on the conifold}
\label{sec:duality}

In this section, we consider the moduli region outside the chamber $C_0$, i.e. away from the singular point $z=0$. We particularly focus on the conifold case. The generalization to the ALE space will be given in section \ref{sec:orbifold}.
When the moduli parameter $z$ is outside $C_0$, the corresponding FI parameter does not satisfy $\theta_1,\theta_2<0$. Then the generating function of the Witten index \eqref{eq:partition0}  should be modified, due to the wall-crossing phenomena. However, it was pointed out in \cite{Chuang:2008aw, Chuang:2009pd, Aganagic:2010qr} that, if we change the basis of the D-brane charge via the Seiberg duality, we recover the condition $\theta_1,\theta_2<0$ in a dual frame. Below, we discuss how the Seiberg duality reproduces the wall-crossing phenomena reviewed in section \ref{sec:WC}.

Let us first recall the Seiberg duality \cite{Seiberg:1994pq} of a quiver gauge theory \cite{Berenstein:2002fi}. We fix a quiver node $i$ and take the duality transformation with respect to $i$. We first introduce a ``meson field''
\begin{eqnarray}
M = X^{(s)}X^{(e)}
\label{eq:meson}
\end{eqnarray}
for each arrow $X^{(s)}$ starting at $i$ and each arrow $X^{(e)}$ ending at $i$. The meson $M$ is expressed as an arrow from the starting node of $X^{(e)}$ to the ending node of $X^{(s)}$. We then reverse all the directions of $X^{(s)}$ and $X^{(e)}$, and relabel them as $\tilde{X}^{(s)}$ and $\tilde{X}^{(e)}$ respectively.  The superpotential of the dual theory is given by
\begin{eqnarray}
\widetilde{W} = W + \sum_{a,b}{\rm tr}(M_{ab} \tilde{X}^{(e)}_b \tilde{X}^{(s)}_a),
\end{eqnarray}
where $a$ and $b$ run over the arrows starting and ending at the node $i$, respectively.
The first term $W$ is the superpotential of the original theory, in which all $X^{(s)}_a$ and $X^{(e)}_b$ are rewritten in terms of $M_{ab}$ through \eqref{eq:meson}. After this operation, $\widetilde{W}$ could include some quadratic terms. If $\widetilde{W}$ contains a quadratic term $X_1X_2$, then we can integrate out $X_1$ and $X_2$ by the F-term conditions
\begin{eqnarray}
\frac{\partial \widetilde{W}}{\partial X_1} = 0,\qquad \frac{\partial \widetilde{W}}{\partial X_2} = 0.
\end{eqnarray}
We denote by $\widetilde{Q}$ the resulting quiver diagram of the dual theory.

Physically, the Seiberg duality changes the basis of the D-brane charges. Suppose that the reference representation $R$ of $Q$ is mapped to a representation $\widetilde{R}$ of $\widetilde{Q}$ by the duality transformation. We denote by $N_k$ the dimension of the vector space associated with the node $k$ in $R$. We similarly denote by $\widetilde{N}_k$ the dimension in $\widetilde{R}$. Then $\{N_k\}$ and $\{\widetilde{N}_k\}$ are related by
\begin{eqnarray}
\widetilde{N}_i = \sum_{j(\neq i)}N_jn_{ji} - N_i,\qquad \widetilde{N}_k = N_k \quad \text{for}\quad k\neq i,
\end{eqnarray}
where $n_{ji}$ is the number of arrows {\it from} the node $i$ {\it to} $j$ in the original quiver $Q$. This relation implies that, when moving from $R$ to $\widetilde{R}$, the unit charges $\gamma^{(k)}$ transform as
\begin{eqnarray}
\gamma^{(i)} \to -\gamma^{(i)},\qquad \gamma^{(k)} \to \gamma^{(k)} + n_{ki}\gamma^{(i)} \quad \text{for}\quad k\neq i.
\label{eq:shift-charge}
\end{eqnarray}
 Correspondingly, the FI parameters $\{\theta_k\}$ transform as
\begin{eqnarray}
\theta_i \to -\theta_i,\qquad \theta_k \to \theta_k + n_{ki}\theta_i \quad \text{for}\quad k\neq i.
\end{eqnarray}
 Note that this maps $\theta(R)=0$ to $\theta(\widetilde{R}) = 0$.

\subsection{The chamber $C_{n\geq 0}$}
\label{subsec:Ckp}

Let us now focus on the conifold case. We start in the chamber $C_0$ and first {\it decrease} ${\rm Im}\,z$ along the red arrow in figure \ref{fig:wall}. For example, after $z$ crosses the wall $W_0^{+1}$ to enter the chamber $C_1$, it now satisfies
\begin{eqnarray}
{\rm Im}\left((z+1)e^{-2i\varphi}\right)<0,\qquad {\rm Im}\left(-ze^{-2i\varphi}\right)<0.
\label{eq:moduli-C1}
\end{eqnarray}
The FI parameters \eqref{eq:FI-K} then satisfy $\theta_1>0$ and $\theta_2<0$.
However, by taking the Seiberg duality, we can recover $\theta_1,\theta_2<0$ in the dual frame. To see this, let us consider the duality transformation with respect to the node $1$, which changes the FI parameters as
\begin{eqnarray}
\theta_1 \to -\theta_1,\qquad \theta_2\to \theta_2 + 2\theta_1,\qquad \theta_* \to \theta_*.
\label{eq:FI2}
\end{eqnarray}
The relation \eqref{eq:FI-K} is now mapped to
\begin{eqnarray}
\theta_1 = \frac{2}{\Lambda^2}{\rm Im}(-ze^{-2i\varphi}),\qquad \theta_2 = \frac{2}{\Lambda^2}{\rm Im}\left((1+z)e^{-2i\varphi}\right).
\label{eq:identification3}
\end{eqnarray}
Since $z$ satisfies \eqref{eq:moduli-C1}, we restore $\theta_1,\theta_2<0$ in this duality frame. We denote this duality frame by $\mathcal{F}_{1}$, and the original frame by $\mathcal{F}_{0}$.
\begin{figure}
\begin{center}
\includegraphics[width=4cm]{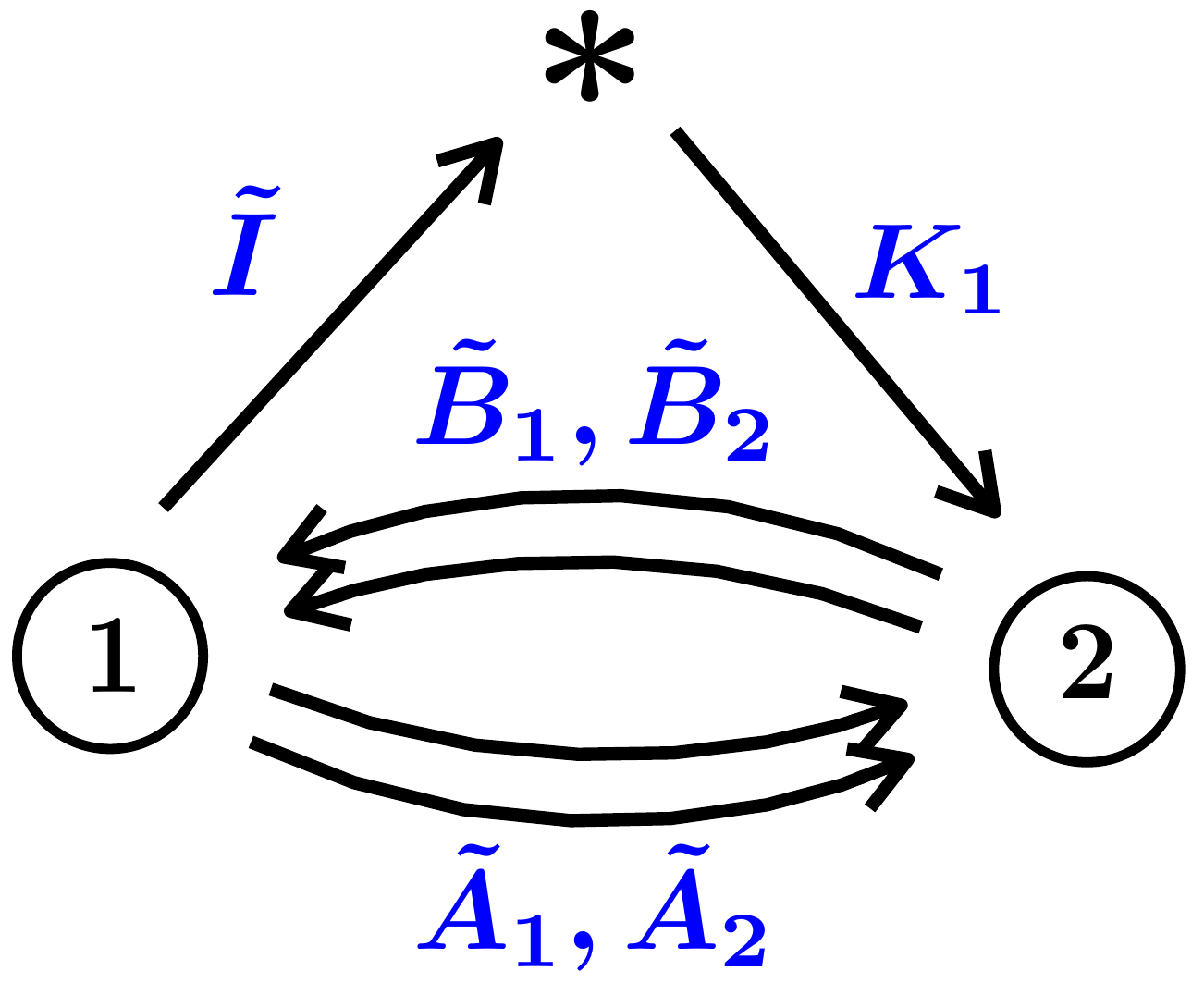}
\caption{The quiver diagram $\widetilde{Q}$ in the dual frame. Note that the roles of the quiver nodes $1$ and $2$ are exchanged.}
\label{fig:dual-quiver1}
\end{center}
\end{figure}

Let us consider the quiver diagram $\widetilde{Q}$ in the dual frame $\mathcal{F}_{1}$. There are four meson fields $M_{ba}$ associated with $B_bA_a$ as well as two meson fields $K_{b}$ associated with $B_{b}I$. We denote the arrows obtained by reversing $A_a,B_b$ and $I$ by $\tilde{A}_a,\tilde{B}_b$ and $\tilde{I}$, respectively. The superpotential in the dual frame is then written as
\begin{eqnarray}
\widetilde{W} = {\rm tr}(M_{12}M_{21}) - {\rm tr}(M_{11}M_{22}) + JK_2 + \sum_{a,b=1,2}{\rm tr}(M_{ba}\tilde{A}_a\tilde{B}_b) + \sum_{b=1,2}{\rm tr}(K_{b}\tilde{I}\tilde{B}_{b}).
\end{eqnarray}
We can integrate out $M_{ba}, K_2$ and $J$ to obtain
\begin{eqnarray}
\widetilde{W} =  {\rm tr}(\tilde{A}_1\tilde{B}_1\tilde{A}_2\tilde{B}_2) -{\rm tr}(\tilde{A}_1\tilde{B}_2\tilde{A}_2\tilde{B}_1) + \tilde{I}\tilde{B}_1K_1.
\label{eq:dual-superpot}
\end{eqnarray}
The quiver diagram $\widetilde{Q}$ is now depicted as in figure \ref{fig:dual-quiver1}.
Note here that this quiver $\widetilde{Q}$ with the potential $\widetilde{W}$ is almost the same as the original quiver $Q$ with $W$ in the frame $\mathcal{F}_{0}$. The only difference is that the roles of the nodes $1$ and $2$ are exchanged. The unit charges associated with the quiver nodes are now
\begin{eqnarray}
\gamma^{(1)} = -\beta,\qquad \gamma^{(2)} = \beta-dV, \qquad \gamma^{(*)} = -\mathcal{D}.
\label{eq:unit1}
\end{eqnarray}
Note that all the central charges $Z(\gamma^{(k)})$ lie in a convex cone because $z$ satisfies \eqref{eq:moduli-C1} and $0<{\rm Re}\,z<1$.

We can generalize this to the chamber $C_{n\geq 2}$, by performing further duality transformations. Let us denote by $T_1$ and $T_2$ the duality transformations with respect to the nodes $1$ and $2$, respectively. We define a chain of $n$ duality transformations by
\begin{eqnarray}
\mathcal{T}_n := \left\{
\begin{array}{l}
(T_2T_1) \cdots (T_2T_1) \quad \text{for}\quad n\in 2\mathbb{N}\\
T_1(T_2T_1) \cdots (T_2T_1) \quad \text{for}\quad n\in 2\mathbb{N}+1\\
\end{array}
\right.,
\end{eqnarray}
where either product has $n$ $T$'s. Suppose that we act the transformation $\mathcal{T}_n$ on the duality frame $\mathcal{F}_0$. We denote the resulting duality frame by $\mathcal{F}_n$. In the frame $\mathcal{F}_n$, the FI parameters are given by
\begin{eqnarray}
\theta_1 = \frac{2}{\Lambda^2}{\rm Im}\left((z +n)e^{-2i\varphi}\right),\qquad \theta_2 = \frac{2}{\Lambda^2}{\rm Im}\left((1-n-z)e^{-2i\varphi}\right)
\label{eq:FI-even}
\end{eqnarray}
if $n\in 2\mathbb{N}$, and by
\begin{eqnarray}
\theta_1 = \frac{2}{\Lambda^2}{\rm Im}\left((1-n-z)e^{-2i\varphi}\right),\qquad \theta_2 = \frac{2}{\Lambda^2}{\rm Im}\left((z +n)e^{-2i\varphi}\right)
\label{eq:FI-odd}
\end{eqnarray}
if $n\in2\mathbb{N}+1$. In either case, the condition $\theta_1,\theta_2<0$ is equivalent to
\begin{eqnarray}
{\rm Im}\left((z+n)e^{-2i\varphi}\right) < 0,\qquad {\rm Im}\left((1-n-z)e^{-2i\varphi}\right)<0,
\end{eqnarray}
which exactly coincides with \eqref{eq:Ck}! Therefore, {\it if $z$ is in the chamber $C_n$ then we have $\theta_1,\theta_2<0$ in the duality frame $\mathcal{F}_n$, and the converse is also true.} Moreover, the quiver diagram in $\mathcal{F}_n$ is the same as in figure \ref{fig:quiver-conifold} if $k\in 2\mathbb{N}$, while it is the same as in figure \ref{fig:dual-quiver1} if $n\in 2\mathbb{N}+1$.
The unit charges in the frame $\mathcal{F}_n$ are given by
\begin{eqnarray}
\gamma^{(1)} = \beta -ndV,\qquad \gamma^{(2)} = -\beta -(1-n)dV,\qquad \gamma^{(*)} = -\mathcal{D}
\label{eq:unit3}
\end{eqnarray}
if $n\in 2\mathbb{N}$, and by
\begin{eqnarray}
\gamma^{(1)} = -\beta - (1-n)dV,\qquad \gamma^{(2)} = \beta -ndV,\qquad \gamma^{(*)} = -\mathcal{D}
\label{eq:unit4}
\end{eqnarray}
if $n\in 2\mathbb{N}+1$. In either case, the central charges $Z(\gamma^{(k)})$ lie in a convex cone if $z$ is in $C_n$ satisfies $0<{\rm Re}\,z<1$. 

Let us now consider the BPS partition function for $z\in C_n$. Since we have $\theta_1,\theta_2<0$ in the duality frame $\mathcal{F}_{n}$, the partition function is given by \eqref{eq:partition0} with a suitable identification of the fugacities. Recall that, if the quiver diagram is given by figure \ref{fig:quiver-conifold}, then the fugacities $x,y$ and $w$ are associated with $\gamma^{(1)},\gamma^{(2)}$ and $\gamma^{(*)}$, respectively. This is the case in the duality frame $\mathcal{F}_{n}$ for $n\in 2\mathbb{N}$. On the other hand, in the duality frame $\mathcal{F}_{n}$ for $n\in2\mathbb{N}+1$, the quiver diagram is given by figure \ref{fig:dual-quiver1}. Since the roles of the nodes $1$ and $2$ are exchanged,  $x,y$ and $w$ are now associated with $\gamma^{(2)},\gamma^{(1)}$ and $\gamma^{(*)}$, respectively. In either case, \eqref{eq:unit3} and \eqref{eq:unit4} imply
\begin{eqnarray}
x = q^nQ,\qquad y = q^{1-n}Q^{-1},\qquad w = 1.
\label{eq:change}
\end{eqnarray}
By substituting this into \eqref{eq:partition0}, the BPS partition function for $z\in C_n$ is evaluated as
\begin{eqnarray}
\mathcal{Z}= \prod_{k=1}^\infty \frac{1}{1-q^k}\prod_{\ell=0}^\infty(1-q^\ell(q^nQ))=\prod_{k=1}^\infty\frac{1}{1-q^k}\prod_{\ell=n}^\infty(1-q^\ell Q).
\end{eqnarray}
This is in perfect agreement with \eqref{eq:formula2} obtained by using the wall-crossing formula. Note that the relation \eqref{eq:change} was already pointed out in \cite{Nishinaka:2011sv}. Here, we have derived it from the Seiberg duality of the quiver theory.

\subsection{The chamber $C_{n<0}$}

We now turn to the chambers $C_{n<0}$. We start with the chamber $C_0$ and {\it increase} ${\rm Im}\,z$ while keeping $0<{\rm Re}\,z<1$. Let us extend the definition of $\mathcal{T}_n$ for $n<0$ by
\begin{eqnarray}
\mathcal{T}_n := \left\{
\begin{array}{l}
(T_1T_2) \cdots (T_1T_2) \quad \text{for}\quad -n \in 2\mathbb{N}\\
T_2(T_1T_2)\cdots (T_1T_2)\quad \text{for}\quad -n\in 2\mathbb{N}+1\\
\end{array}
\right.,
\label{eq:prod1}
\end{eqnarray}
where either product has $|n|$ $T$'s.
We define the duality frame $\mathcal{F}_n$ for $n<0$ by performing $\mathcal{T}_n$ in $\mathcal{F}_0$. In the duality frame $\mathcal{F}_{n<0}$, the FI parameters are given by \eqref{eq:FI-even} if $n\in 2\mathbb{Z}$, while it is given by \eqref{eq:FI-odd} if $n\in2\mathbb{Z}+1$. In either case, the condition $\theta_1,\theta_2<0$ in the duality frame $\mathcal{F}_n$ is equivalent to $z\in C_{n}$. The quiver diagram in the duality frame $\mathcal{F}_{n<0}$ is the same as in figure \ref{fig:quiver-conifold} if $-n\in 2\mathbb{N}$ while it is the same as in figure \ref{fig:dual-quiver1} if $-n\in 2\mathbb{N}+1$. The corresponding superpotential is the same as in subsection \ref{subsec:Ckp}.

An essential difference from $\mathcal{F}_{n\geq 0}$ appears in the unit charges associated with the quiver nodes. Let us first consider the duality frame $\mathcal{F}_{-1}$. The relation \eqref{eq:shift-charge} now implies that
\begin{eqnarray}
\gamma^{(1)} = -\beta -2dV,\qquad \gamma^{(2)}= \beta + dV,\qquad \gamma^{(*)} = -\mathcal{D} -\beta -dV,
\label{eq:unit2}
\end{eqnarray}
in the frame $\mathcal{F}_{-1}$.
Note here that the unit charge $\gamma^{(*)}$ is also shifted because there is an arrow $J$ from $2$ to $*$ in the quiver in figure \ref{fig:quiver-conifold}. This is quite different from \eqref{eq:unit1} for $\mathcal{F}_{n\geq 0}$. 
In the duality frame $\mathcal{F}_{n<0}$, the unit charges $\gamma^{(k)}$ are written as
\begin{eqnarray}
\gamma^{(1)} = \beta + |n|dV,\quad \gamma^{(2)} = -\beta -(|n|+1)dV,\quad \gamma^{(*)} = -\mathcal{D}-|n|\beta - \frac{|n|(|n|+1)}{2}dV,\qquad
\label{eq:unit5}
\end{eqnarray}
if $-n\in 2\mathbb{N}$, while they are written as
\begin{eqnarray}
\gamma^{(1)} = -\beta - (|n|+1)dV,\quad \gamma^{(2)} = \beta +|n|dV,\quad \gamma^{(*)} = -\mathcal{D} - |n|\beta - \frac{|n|(|n|+1)}{2}dV.\qquad
\label{eq:unit6}
\end{eqnarray} 
if $-n\in 2\mathbb{N}+1$. Thus, $\gamma^{(*)}$ always carries some D2-D0 charges in the frame $\mathcal{F}_{n<0}$.
 Note that the central charges $Z(\gamma^{(k)})$ lie in a convex cone if $z$ is in $C_n$ and satisfies $0<{\rm Re}\,z<1$. 

Let us now consider the BPS partition function for $z\in C_{n<0}$. 
Since we have $\theta_1,\theta_2<0$ in the duality frame $\mathcal{F}_n$, the partition function is again given by \eqref{eq:partition0} with a suitable identification of the fugacities. The unit charges \eqref{eq:unit5} and \eqref{eq:unit6} now imply that the fugacities are identified as
\begin{eqnarray}
x = q^{-|n|}Q,\qquad y = q^{|n|+1}Q^{-1},\qquad w = q^{\frac{|n|(|n|+1)}{2}}Q^{-|n|}.
\label{eq:change2}
\end{eqnarray}
Note that, since $\gamma^{(*)}$ carries some D2-D0 charge, we here have $w\neq 1$. By substituting \eqref{eq:change2} into \eqref{eq:partition0}, the BPS partition function for $z\in C_{n<0}$ is evaluated as
\begin{eqnarray}
\mathcal{Z}&=&q^{\frac{|n|(|n|+1)}{2}}Q^{-|n|}\prod_{k=1}^\infty\frac{1}{1-q^k}\prod_{\ell=0}^\infty(1-q^\ell(q^{-|n|}Q))
\nonumber\\
&=& (-1)^{|n|}\prod_{k=1}^\infty\frac{1}{1-q^k}\prod_{\ell=0}^\infty(1-q^\ell Q)\prod_{m=1}^{|n|}(1-q^{m}Q^{-1}).
\end{eqnarray}
Note that, up to the overall sign, this agrees with \eqref{eq:formula} obtained by using the wall-crossing formula! The overall sign factor just changes whether a bosonic state contributes $1$ or $-1$ to the Witten index. We stress here that the relation \eqref{eq:change2} was not studied in \cite{Nishinaka:2011sv}, and has been newly discovered here.

\section{Interpretation in the brane tiling}
\label{sec:dimer}

\begin{figure}
\begin{center}
\includegraphics[width=3.8cm]{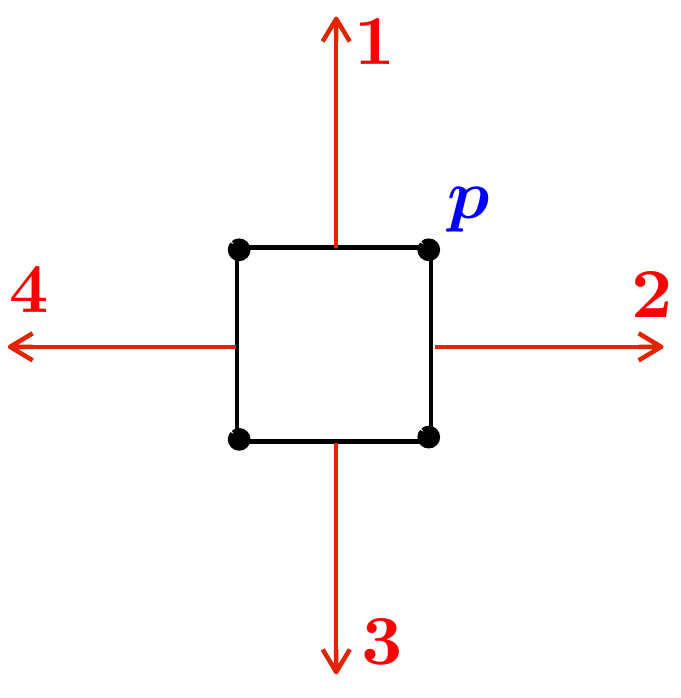}\qquad\qquad\qquad
\includegraphics[width=3.8cm]{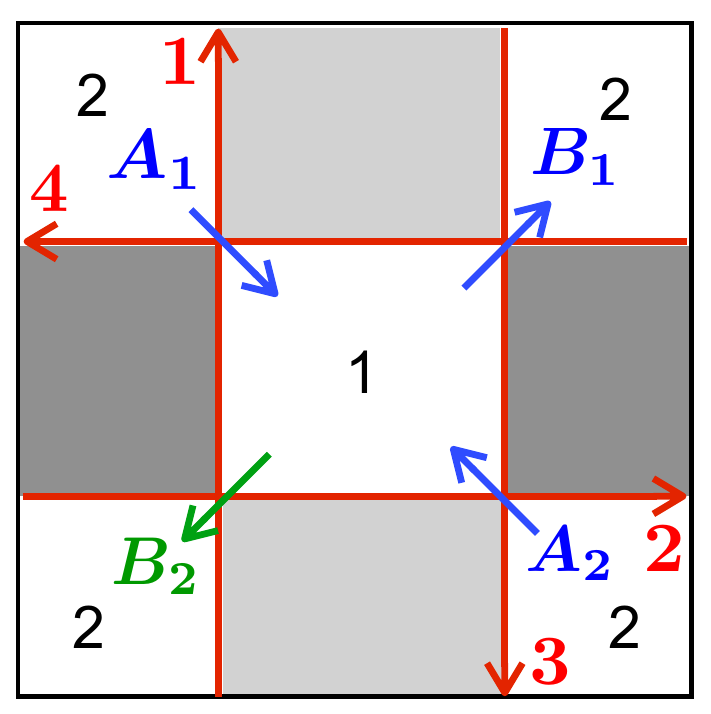}
\caption{Left: The toric diagram of the conifold. \; Right: The corresponding brane tiling system. The red arrows stand for NS5-branes, while the blue and green arrows express chiral multiplets.  When we put a D4-brane on a divisor associated with $p$, the corresponding NS5-branes $\ell_1,\ell_2$ are identified with the first and second red arrows. Since $B_2$ is located at their intersection, we identify $X_F=B_2$. Therefore, in this case, $\mathfrak{i}$ and $\mathfrak{j}$ are the white tiles $1$ and $2$, respectively.}
\label{fig:toric}
\end{center}
\end{figure}

We here give an interpretation of the Seiberg duality in the brane tiling associated with the D4-D2-D0 states. The brane tiling is a technique to read off the quiver diagram $Q$ and superpotential $W$ \cite{Hanany:2005ve, Franco:2005rj, Franco:2005sm, Hanany:2005ss} (See \cite{Kennaway:2007tq, Yamazaki:2008bt} for reviews.). We briefly review the technique in the next three paragraphs, following \cite{Ooguri:2008yb}.

Let $\Sigma$ be the toric diagram of the Calabi-Yau singularity $Y_\Sigma$. By definition, $\Sigma$ is a convex lattice polygon (figure \ref{fig:toric}). We denote the set of lattice points in $\Sigma$ by $\Sigma_0$, and the set of line segments in $\Sigma$ by $\Sigma_1$.   The topology of the Calabi-Yau singularity is completely determined by $\Sigma$. Since it is toric, $Y_\Sigma$ admits a natural torus action. By taking the T-duality transformation in the two directions of the torus action, we can map the Calabi-Yau singularity $Y_\Sigma$ to intersecting NS5-branes in flat spacetime. For each line segment $s\in \Sigma_1$, we have a single NS5-brane wrapped on a one-cycle $\mathcal{C}_s$ of the torus. The winding number of $\mathcal{C}_s$ coincides with the slope of the out-going normal of the corresponding line segment $s$ (figure \ref{fig:toric}). The intersecting NS5-branes divide the torus into several ``tiles.'' Some of the tiles are assigned definite orientations by the NS5-branes. The conservation of the NS5-charge implies that the NS5-branes fill up all such tiles \cite{Imamura:2006ub}. We assign the color of light gray or dark gray to each such tile, depending on the orientation of the NS5-branes. The other tiles are assigned white color (figure \ref{fig:toric}).

Suppose that we originally have $N_0$ D0-branes and $N_2$ D2-branes at the Calabi-Yau singularity $Y_\Sigma$. The T-duality transformation maps them into D2-branes wrapping on $T^2$. The $N_0$ D2-branes are wrapped on the whole $T^2$ while the $N_2$ D2-branes are suspended in one of the white tiles. Then each white tile gives a gauge multiplet. The rank of the gauge group depends on how many D2-branes exist in the white tile. At each intersection of the white tiles, there is a chiral multiplet expressed as an arrow. The direction of the arrow is determined by the relative positions of the adjacent dark and light gray tiles. The superpotential for the chiral multiplets comes from the gray tiles. Namely, if a gray tile is surrounded by a chain of chiral multiplets $X_1,\cdots, X_n$ then we have
\begin{eqnarray}
\pm{\rm tr}(X_1\cdots X_n)
\end{eqnarray}
in the superpotential. Here the sign depends on whether the tile is dark-gray or light-gray colored.

\begin{figure}
\begin{center}
\includegraphics[width=3.5cm]{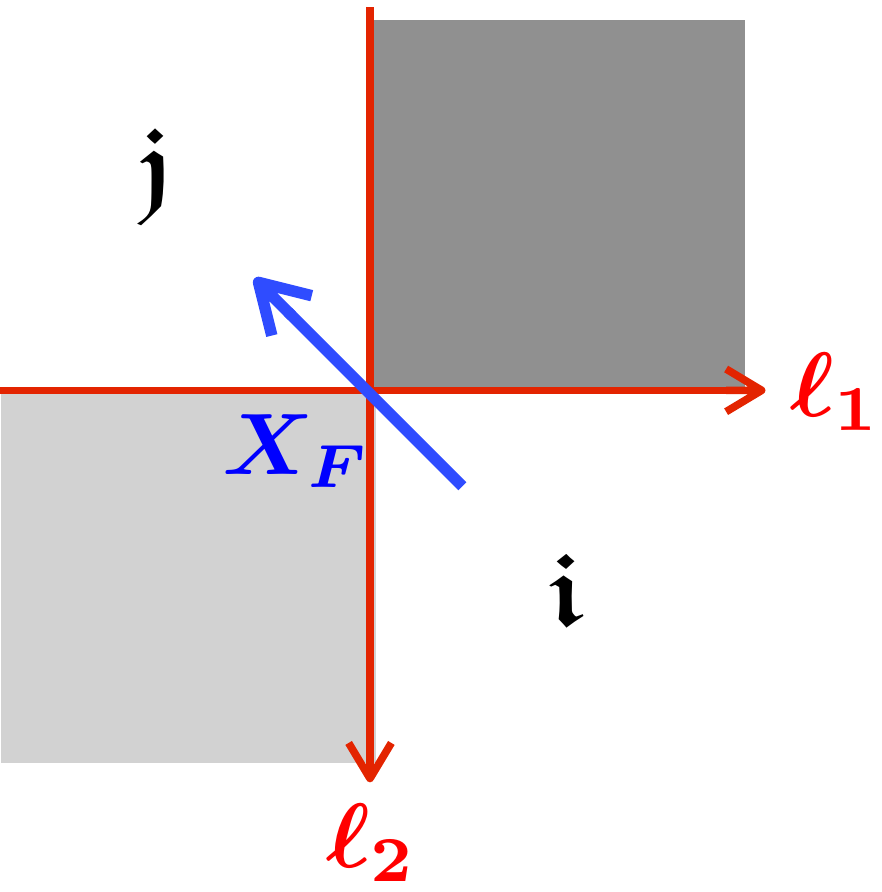}\qquad\qquad\qquad
\includegraphics[width=3.7cm]{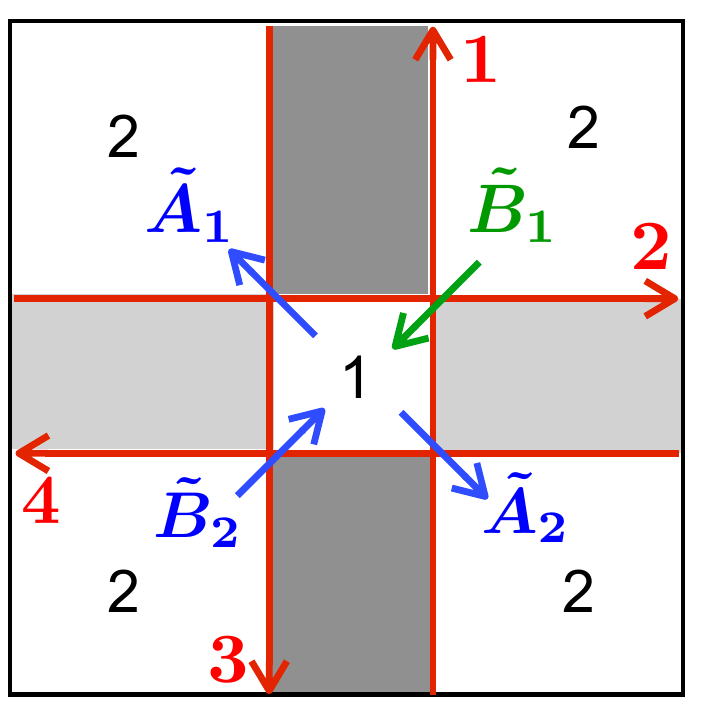}
\caption{Left: The intersection of $\ell_1$ and $\ell_2$ is attached to two dynamical D2-branes $\mathfrak{i}$ and $\mathfrak{j}$. \; Right: By moving the NS5-branes, we can flip the orientation of the D2-branes filling up a white tile.  Here, we start with the configuration in figure 6, and shrink the white tile $1$, and spread it out with the opposite orientation. The directions of the chiral multiplets are reversed. We now identify $X_F = \widetilde{B}_1$.}
\label{fig:flip_orientation}
\end{center}
\end{figure}

Let us now consider an additional D4-brane wrapped on a toric divisor $D$ of $Y_\Sigma$ \cite{Nishinaka:2013}. Such a divisor is associated with a lattice point $p\in \Sigma_0$. We assume $p$ is located at a {\it corner} of the diagram $\Sigma$ so that $D$ is not degenerate in the singular Calabi-Yau limit of $Y_{\Sigma}$. Such a lattice point $p$ is attached to two line segments in $\Sigma_1$. The line segments are associated with two NS5-branes $\ell_1$ and $\ell_2$ in the brane tiling. Since $\ell_1$ and $\ell_2$ are not parallel, they intersect with each other in the torus. We pick one of the intersections and denote it by $*$.\footnote{If $\ell_1$ and $\ell_2$ intersect at several points in $T^2$, the choice of the location of $*$ is related to the holonomy of the gauge field at infinity on the D4-brane \cite{Nishinaka:2013}.} We use the same symbol as the D4-node here, because the D4-brane is localized at this point after the T-duality. To be more precise, the T-duality maps the D4-brane to a non-compact D2-brane bounded by the two NS5-branes $\ell_1, \ell_2$ \cite{Nishinaka:2013}. Since the D4-brane originally wraps on the whole $T^2$, the non-compact D2-brane is localized at a point in $T^2$. This point should be the intersection of $\ell_1$ and $\ell_2$ because the D2-brane is bounded by $\ell_1,\ell_2$. Note that the intersection point $*$ is always attached to two white tiles. We denote them by $\mathfrak{i}$ and $\mathfrak{j}$ so that there is a chiral multiplet $X_F$ from $\mathfrak{i}$ to $\mathfrak{j}$ (figure \ref{fig:flip_orientation}). Then there is a ``quark'' $I$ from $*$ to $\mathfrak{i}$ as well as an ``anti-quark'' $J$ from $\mathfrak{j}$ to $*$. The D4-node $*$ induces an additional superpotential term
\begin{eqnarray}
JX_FI.
\end{eqnarray}
Thus, we can derive the quiver diagram $Q$ and superpotential $W$ for the D4-D2-D0 states from the corresponding brane tiling.

We are now ready to see an interpretation of the Seiberg duality in the brane tiling. Without the D4-brane, such an interpretation is well-known \cite{Franco:2005rj}. Suppose that we take the duality transformation with respect to a quiver node $i\neq *$. The node $i$ is associated with a white tile $f_i$ in the brane tiling. We want to change the basis of the unit charge as in \eqref{eq:shift-charge}. In particular, a D2-brane suspended in the white tile $f_i$ should be mapped to an {\it anti} D2-brane. This is realized by moving the NS5-branes around $f_i$ so that the orientation of the D2-brane in $f_i$ flips (figure \ref{fig:flip_orientation}). In this process, the white tile $f_i$ shrinks once and spreads again with the opposite orientation. From the resulting brane configuration, we can read off the quiver and superpotential in the Seiberg dual frame.

Let us now consider where the D4-node $*$ is located after the duality transformation. Recall that $*$ should be located at the intersection of the two NS5-branes $\ell_1$ and $\ell_2$. Since the duality transformation deforms the NS5-branes, they could intersect at a different point after the transformation. In the conifold example of figure \ref{fig:toric}, the NS5-branes $\ell_1$ and $\ell_2$ intersect at $B_2$ before the duality transformation. On the other hand, they intersect at $\tilde{B}_1$ after the transformation (figure \ref{fig:flip_orientation}). This implies that the Seiberg duality replaced the superpotential term $JB_2I$ with
\begin{eqnarray}
J\tilde{B}_1I.
\end{eqnarray}
When we rewrite $I$ and $J$ into $K_1$ and $\tilde{I}$ respectively, this procedure perfectly reproduces the superpotential \eqref{eq:dual-superpot} of the dual quiver $\widetilde{Q}$. Thus, the location of the D4-node is always read off from the intersection of the NS5-branes $\ell_1,\ell_2$.

\section{Application to the orbifold $\mathbb{C}^2/\mathbb{Z}_2$}
\label{sec:orbifold}

We now consider the Seiberg duality of the quiver theory on the D4-D2-D0 states on $\mathbb{C}^2/\mathbb{Z}_2$. One subtlety is that the quiver in figure \ref{fig:quiver-orbifold} now contains adjoint matters. The Seiberg duality with adjoint matters non-trivially depends on the form of the superpotential \cite{Kutasov:1995ve, Berenstein:2002fi}, which makes the duality transformation more involved. However, for quivers associated with the D4-D2-D0 states, we can easily identify the dual theory following the prescription described in the previous section. 

\begin{figure}
\begin{center}
\includegraphics[width=2.7cm]{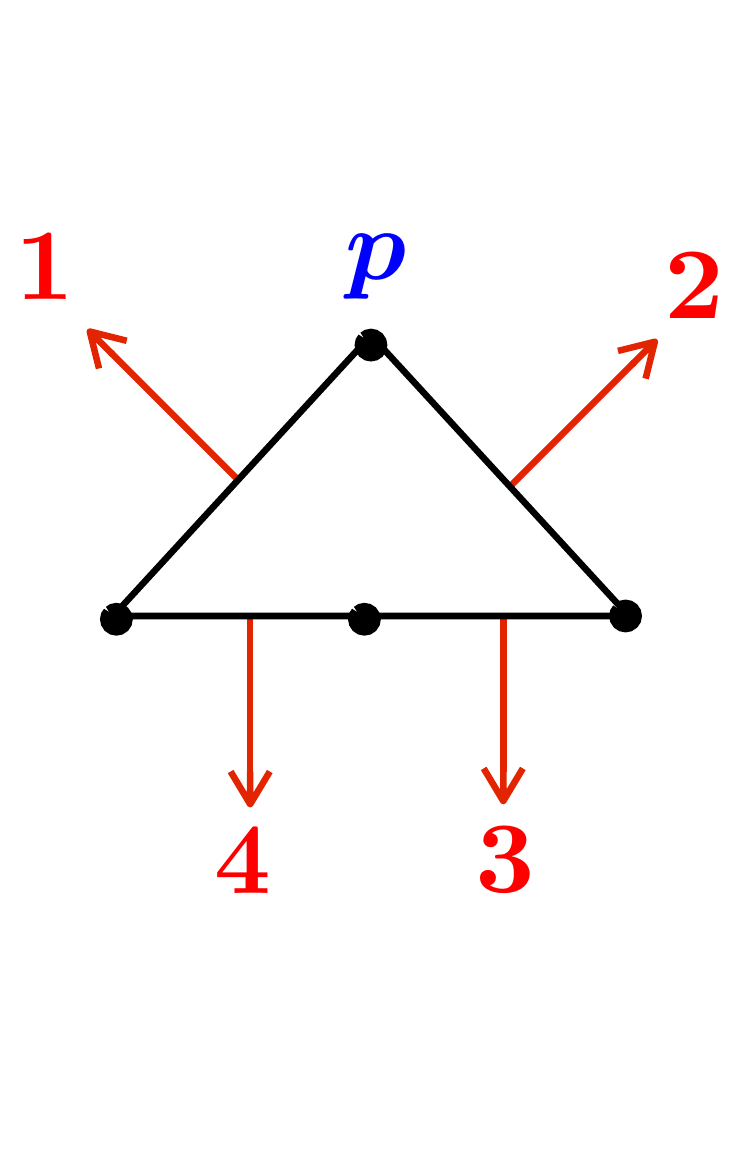}\qquad\qquad
\includegraphics[width=4cm]{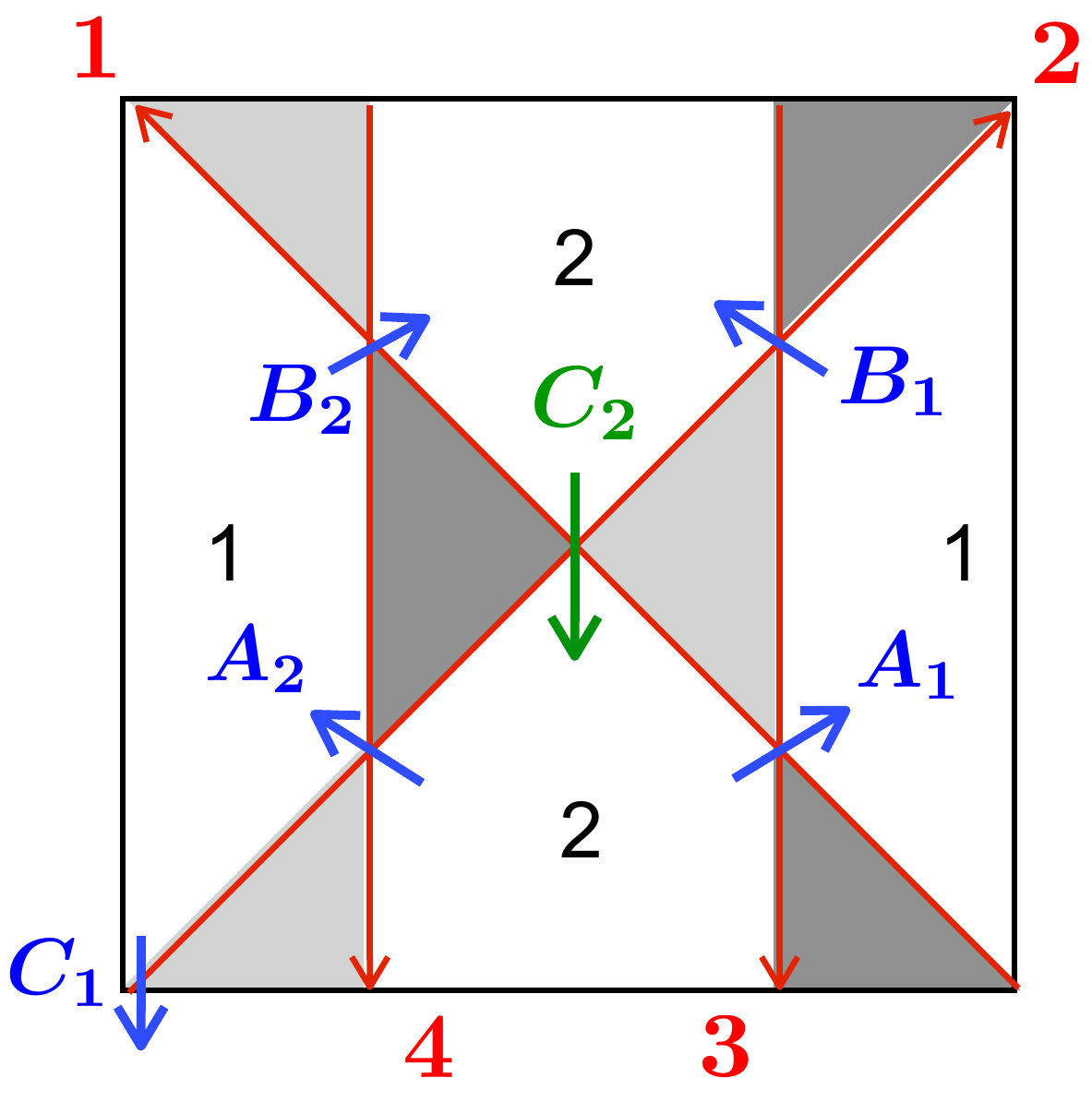}\qquad\qquad
\includegraphics[width=4.1cm]{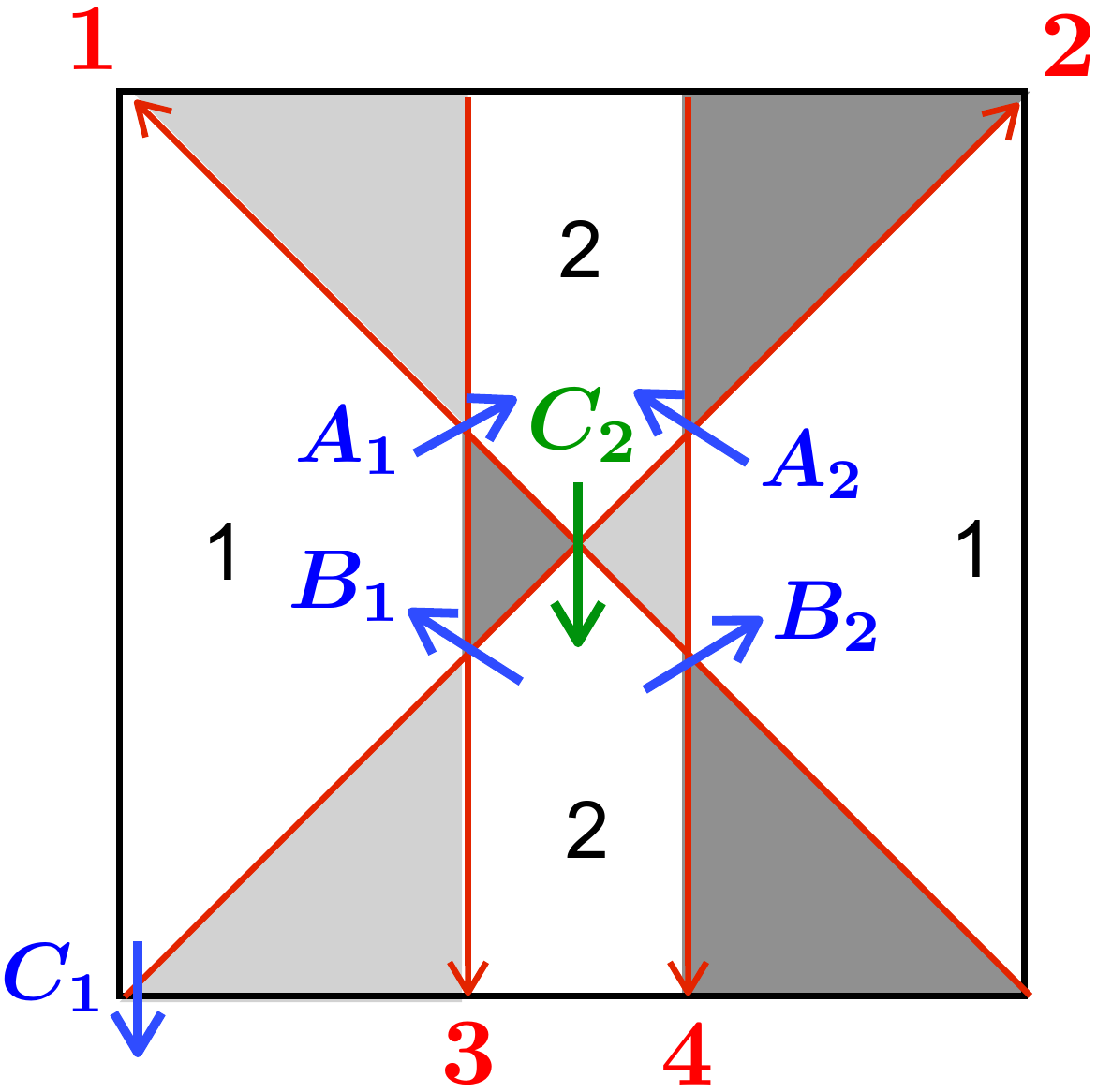}
\caption{Left: The toric diagram of the orbifold $\mathbb{C}^2/\mathbb{Z}_2$. The D4-brane is on the divisor associated with $p$. \; Middle: The brane tiling associated with D4-D2-D0 states on the orbifold. Here $\ell_1,\ell_2$ are identified with the first and second NS5-branes. The corresponding quiver diagram is shown in figure 3. \; Right: The Seiberg duality with respect to the node $2$ is realized by crossing the third and fourth NS5-branes, which does not change the tiling structure. Moreover, the duality does not move the intersection points of $\ell_1,\ell_2$.}
\label{fig:dimer-orbifold}
\end{center}
\end{figure}

We first divide the moduli space into chambers $C_{n}$ defined by \eqref{eq:Ck}. Since $z$ is now restricted to the upper half plane, we only consider $n\leq 0$ in this section. As described in section \ref{sec:duality}, if $z$ is in the chamber $C_{n}$ then $\theta_1,\theta_2<0$ follows in the duality frame $\mathcal{F}_n$. Below, we identify the quiver and superpotential in the duality frame $\mathcal{F}_{n\leq 0}$.
In the frame $\mathcal{F}_0$, the quiver is given by figure \ref{fig:quiver-orbifold} and the superpotential is written as \eqref{eq:superpot-orbifold}. The corresponding brane tiling is shown in the middle picture of figure \ref{fig:dimer-orbifold}. In particular, the NS5-branes $\ell_1$ and $\ell_2$ are here the first and second NS5-branes. 
Let us now move to the duality frame $\mathcal{F}_{-1}$. The Seiberg duality from $\mathcal{F}_{0}$ to $\mathcal{F}_{-1}$ reverses the orientation of the D2-branes in the white tile $2$. This is realized by crossing the third and fourth NS5-branes. The resulting brane tiling is shown in the right picture of figure \ref{fig:dimer-orbifold}. Note that, although the basis of the D-brane charge has been changed, the tiling structure is exactly the same as in the middle picture of figure \ref{fig:dimer-orbifold}. Moreover, this procedure does not move the intersection point of the NS5-branes $\ell_1,\ell_2$. This means that {\it the duality frame $\mathcal{F}_{-1}$ has the same quiver and superpotential as the frame $\mathcal{F}_{0}$.} Note that, unlike the conifold case, the role of the nodes $1$ and $2$ are not exchanged. The only difference from the duality frame $\mathcal{F}_0$ is that the unit charges are now given by \eqref{eq:unit2}.
In general, we can move from $\mathcal{F}_{0}$ to $\mathcal{F}_{n<0}$ by taking the duality transformation \eqref{eq:prod1}. The quiver and superpotential are always invariant under the duality transformations. The unit charges in the frame $\mathcal{F}_{n}$ are given by
\begin{eqnarray}
\gamma^{(1)} = \beta -ndV,\quad \gamma^{(2)} = -\beta - (1-n)dV,\quad \gamma^{(*)}= -\mathcal{D} + \frac{n}{2}\beta -\frac{n^2}{4}dV
\label{eq:unit-orbifold1}
\end{eqnarray}
if $-n\in 2\mathbb{N}$, while they are written as
\begin{eqnarray}
\gamma^{(1)} = -\beta - (1-n)dV,\quad \gamma^{(2)} = \beta - ndV,\quad \gamma^{(*)}=-\mathcal{D} +\frac{n-1}{2}\beta -\frac{(n-1)^2}{4}dV
\label{eq:unit-orbifold2}
\end{eqnarray}
if $-n\in2\mathbb{N}+1$.

Let us consider the partition function in the chamber $C_{n\leq 0}$. Since the FI parameters satisfy $\theta_1,\theta_2<0$ in the duality frame $\mathcal{F}_{n}$, the partition function is written as \eqref{eq:partition2} with a suitable identification of the fugacities. Recall that $x,y$ and $w$ are fugacities associated with the nodes $1,2$ and $*$, respectively. The unit charges \eqref{eq:unit-orbifold1} and \eqref{eq:unit-orbifold2} then imply that, in the duality frame $\mathcal{F}_{n}$, the fugacities are identified as
\begin{eqnarray}
x = q^n Q,\qquad y = q^{1-n}Q^{-1},\qquad w=q^{\frac{n^2}{4}}Q^{\frac{n}{2}}
\end{eqnarray}
if $-n\in2\mathbb{N}$, and
\begin{eqnarray}
x = q^{1-n}Q^{-1},\qquad y = q^{n}Q,\qquad w= q^{\frac{(n-1)^2}{4}}Q^{\frac{n-1}{2}}
\end{eqnarray}
if $-n\in2\mathbb{N}+1$.
By substituting these into \eqref{eq:partition2}, the partition function is written as
\begin{eqnarray}
\mathcal{Z}
=  \prod_{k=1}^\infty \frac{1}{1-q^k}\sum_{\ell\in\mathbb{Z}}q^{(\ell+\frac{n}{2})^2}Q^{\ell+\frac{n}{2}}
\end{eqnarray}
if $-n\in 2\mathbb{N}$, while it is written as
\begin{eqnarray}
\mathcal{Z}
=  \prod_{k=1}^\infty \frac{1}{1-q^k}\sum_{\ell\in\mathbb{Z}}q^{(-\ell-\frac{1-n}{2})^2}Q^{-\ell-\frac{1-n}{2}}
\end{eqnarray}
if $-n\in2\mathbb{N}+1$. In either case, by relabeling the index $\ell$ in the summation, we find that the partition function $\mathcal{Z}$ coincides with \eqref{eq:affine}! Therefore, the BPS partition function is exactly constant in the whole moduli space, and always given by the character of $\widehat{su}(2)_{1}$. This is in perfect agreement with the wall-crossing formula reviewed in subsection \ref{subsec:orbifold}.

\section{Discussion}
\label{sec:discussion}

In this paper, we have studied the wall-crossing phenomena of D4-D2-D0 states on the conifold and orbifold $\mathbb{C}^2/\mathbb{Z}_2$. The K\"ahler moduli dependence of the BPS index is translated into the FI parameter dependence of the Witten index. In section \ref{sec:relation}, we have identified the moduli region in which the quiver quantum mechanics description is reliable. Since such a region depends on the duality frame, we have fixed the duality frame $\mathcal{F}^{(0)}$. In section \ref{sec:duality}, we have shown that the wall-crossings of the D4-D2-D0 states on the resolved conifold are related to the Seiberg dualities. In particular, we have shown that $\theta_1,\theta_2<0$ holds in the duality frame $\mathcal{F}^{(n)}$ if the moduli are in the chamber $C_n$. The generating function of the Witten index is consistent with the wall-crossing formula. In section \ref{sec:dimer}, we have given an interpretation of the Seiberg duality with a D4-node in the brane tiling. In particular, the D4-node is always located at the intersection of the boundary NS5-branes $\ell_1$ and $\ell_2$. In section \ref{sec:orbifold}, we have applied our interpretation to the D4-D2-D0 states on the orbifold $\mathbb{C}^2/\mathbb{Z}_2$. The resulting partition function is always given by the character of $\widehat{su}(2)_1$, which is consistent with the wall-crossing formula.

An interesting observation is that, the quiver diagram for our D4-D2-D0 state has the same form in every Seiberg duality frame. This is quite different from the D6-D2-D0 case studied in \cite{Chuang:2008aw, Chuang:2009pd}. It would be interesting to study the physical origin of this result further. We here stress that the existence of the ``anti-quark'' $J$ is crucial for this result; it kills the unwanted meson field. In fact, if $J$ does not exist, the Seiberg duality leads to the duality cascade for the D6-D2-D0 states. In this sense, $J$ plays an important role in describing the correct D4-D2-D0 wall-crossings.

An interesting future direction will be to study the relation to the $q$-deformed Yang-Mills theory. In the large radius limit, our D4-D2-D0 states are described by the two-dimensional $q$-deformed Yang-Mills theory on $\mathbb{P}^1$ \cite{Aganagic:2004js, Aganagic:2005wn, Griguolo:2006kp}. It would be interesting to study how to describe the wall-crossing phenomena of the D4-D2-D0 states in terms of the $q$-deformed Yang-Mills theory.

\section*{Acknowledgments}

The author would like to thank Emanuel Diaconescu, Greg Moore, So Okada, Satoshi Yamaguchi, Masahito Yamazaki and Yutaka Yoshida for illuminating discussions. He also thanks Marie Nishinaka for important comments on figures in this paper. The work of T.N. is supported in part by the U.S. Department of Energy under grant DE-FG02-96ER40959. 

\appendix

\section{The BPS index}
\label{app:index}

We here review the basic property of the BPS index in $d=4,\mathcal{N}=2$ theories.
The BPS index for an electro-magnetic charge $\gamma\neq 0$ is defined by\footnote{If the theory has $\mathcal{N}>2$ supersymmetry, we should modify this expression to factor out more fermion zero modes.}
\begin{eqnarray}
\Omega(\gamma) = -\frac{1}{2}{\rm Tr}_{\mathcal{H}_\gamma}[(-1)^{F}(2J)^2],
\label{eq:index}
\end{eqnarray}
where the trace is taken over the space $\mathcal{H}_\gamma$ of all {\em one-particle states} carrying charge $\gamma$. We assume the central charge $Z_\gamma$ for $\gamma$ does not vanish, which implies all the states in $\mathcal{H}_{\gamma}$ are massive. The operator $J$ is the third component of the angular momentum operator and $F=2J$ is the fermion number operator. In $d=4,\mathcal{N}=2$ theories, any massive one-particle state belongs to a short or long supersymmetry multiplet. A short multiplet is also called a BPS multiplet, and breaks half the supersymmetry. An irreducible short multiplet has the spin structure
\begin{eqnarray}
 [j]\otimes \left(2[0] \oplus \left[\frac{1}{2}\right]\right),
\label{eq:short}
\end{eqnarray}
where $[j]$ is a spin multiplet of spin $j$. The factor $2[0] \oplus [\frac{1}{2}]$ is called a half-hypermultiplet and formed by the four broken supercharges. On the other hand, a long supersymmetry multiplet breaks all the supersymmetry and has the spin structure
\begin{eqnarray}
[j] \otimes \left(2[0]\oplus \left[\frac{1}{2}\right]\right)\otimes \left(2[0]\oplus \left[\frac{1}{2}\right]\right).
\label{eq:long}
\end{eqnarray}
Here the two half-hypermultiplets are formed by eight broken supercharges. We call a state in a short (long) multiplet a ``BPS (non-BPS) state.''
Since any state in $\mathcal{H}_{\gamma}$ belongs to either \eqref{eq:short} or \eqref{eq:long}, we can factor out the trace over one half-hypermultiplet in \eqref{eq:index} to obtain
\begin{eqnarray}
\Omega(\gamma) = {\rm Tr}_{\mathcal{H}_\gamma/(2[0]\oplus[\frac{1}{2}])}(-1)^{F}.
\end{eqnarray}
This means that only BPS multiplets contribute to $\Omega(\gamma)$.

The BPS index $\Omega(\gamma)$, by definition, depends on the Hilbert space of BPS states with charge $\gamma$. If all the BPS states are stable against the variations of vacuum moduli parameters, then the BPS index is exactly constant in the moduli space. However, in general, the Hilbert space of BPS states {\it does} depend on the vacuum moduli; a BPS state could be unstable if the moduli parameters are tuned. The BPS index $\Omega(\gamma)$ is therefore a non-trivial function on the moduli space. Since $\Omega(\gamma)$ is integer-valued, it can change only discontinuously. A discontinuous change of $\Omega(\gamma)$ under a variation of the moduli is called a ``wall-crossing phenomenon.''

Let us consider when a wall-crossing phenomenon occurs. First of all, any wall-crossing is associated with the appearance or disappearance of some BPS state in the one-particle spectrum. If we tune the moduli parameters so that a BPS state with charge $\gamma$ appear or disappear in the spectrum, the BPS state becomes marginally unstable against decay into other states. Suppose that a BPS state with charge $\gamma$ marginally decays into $n$ one-particle states with charge $\gamma_1,\cdots,\gamma_n$. The charge conservation implies $Z(\gamma) = Z(\gamma_1)  + \cdots + Z(\gamma_n)$, which particularly means
\begin{eqnarray}
|Z(\gamma)|\leq |Z(\gamma_1)|+ \cdots + |Z(\gamma_n)|.
\label{eq:wall}
\end{eqnarray}
On the other hand, the energy-momentum conservation requires
\begin{eqnarray}
|Z(\gamma)| \geq M_1 + \cdots + M_n,
\end{eqnarray}
where $M_i$ is the mass of the $i$-th final state. Because of the the BPS bound $M_i\geq |Z(\gamma_i)|$ for each $i$, this particularly implies 
\begin{eqnarray}
|Z(\gamma)|\geq |Z(\gamma_1)|+ \cdots + |Z(\gamma_n)|.
\label{eq:wall1}
\end{eqnarray}
Given the electro-magnetic charge conservation $\gamma = \gamma_1 + \cdots + \gamma_n$, the conditions \eqref{eq:wall} and \eqref{eq:wall1} are simultaneously satisfied if and only if 
\begin{eqnarray}
{\rm arg}\,Z(\gamma_1) = \cdots = {\rm arg}\,Z(\gamma_n).
\label{eq:wall2}
\end{eqnarray}

Since the central charge $Z(\gamma_i)$ implicitly depends on the vacuum moduli, we can solve this equation for the moduli parameters. The solution gives a subspace of the moduli space. Any wall-crossing phenomenon occur only if the moduli parameters are in such a subspace. Note here that the solution space of \eqref{eq:wall2} is included in that of 
\begin{eqnarray}
 {\rm arg}\,Z(\gamma_1) = {\rm arg}\,Z(\gamma_2 + \cdots+ \gamma_n).
\end{eqnarray}
Therefore, any wall-crossing phenomenon occurs only if there is a two-body decay channel of the form $\gamma\to \gamma_1 + \gamma_2$ such that
\begin{eqnarray}
{\rm arg}\,Z(\gamma_1) = {\rm arg}\, Z(\gamma_2).
\label{eq:wall4}
\end{eqnarray}

\bibliography{ref}

\end{document}